\documentclass[12pt]{amsart}
\usepackage{array,longtable}
\usepackage{bbm}
\usepackage{amssymb,latexsym,fancyhdr,color,graphics}
\usepackage{amsmath, amsthm}
\usepackage{setspace}
\usepackage[dvips]{graphicx}
\usepackage[pdftex]{hyperref}
\usepackage[parfill]{parskip}
\newtheorem{lem}{\bf Lemma}
\newtheorem{asm}{\bf Assumption}
\newcommand{\pff}{\mbox{\bf Proof of Proposition 1:}\hspace{3mm}}
\newtheorem{prop}{\bf Proposition}
\newtheorem{thm}{\bf Theorem}

\newtheorem{rem}{\bf Remark}

\newcommand{\pf}{\mbox{\bf Proof:}\hspace{3mm}}

\definecolor{titlepagegray}{gray}{0.8}

\begin{document}

\title[Utility maximisation and utility indifference price]{Utility maximisation and utility indifference price for exponential semi-martingale models with random factor}
\maketitle

\begin{center}
{\large A. Ellanskaya}\footnote{$^{,2}$ LAREMA, D\'epartement de
Math\'ematiques, Universit\'e d'Angers, 2, Bd Lavoisier - 49045,
{\sc Angers Cedex 01.}\\
$^1$E-mail: Anastasia.Ellanskaya@univ-angers.fr\\
$^2$E-mail: Lioudmila.Vostrikova@univ-angers.fr}{\large
\,\, and  L. Vostrikova$^2$}
\end{center} 
\vspace{0.2in}

\begin{abstract}We consider utility maximization problem for semi-martingale models depending on a random factor $\xi$. We reduce initial maximization problem to the conditional one, given $\xi=u$,  which we solve using dual approach. For HARA utilities we consider information quantities  like Kullback-Leibler information and Hellinger integrals, and corresponding information processes. As a particular case we study
exponential Levy models depending on random factor. In that case the information processes are deterministic and this fact simplify very much indifference price calculus. Then we give the equations for indifference prices. We show that indifference price for seller and minus indifference price for buyer are risk measures. Finally, we apply the results to Geometric Brownian motion case. Using identity in law technique we give the explicit expression for information quantities. Then, the previous formulas for indifference price can be applied.
\end{abstract}
%%%%%%%%%%%%%%%%%%%%%%%%%%%%%%%%%%%%%%%%%%%%%%%%%%%%%%%%%%%%%%%%%%%%%%%%%%%%%%
\begin{section}{Introduction}\label{s0}%%%%%%%%%%%%%%%%%%%%%%%%%%%%%%%%%%%%%%%%
%%%%%%%%%%%%%%%%%%%%%%%%%%%%%%%%%%%%%%%%%%%%%%%%%%%%%%%%%%%%%%%%%%%%%%%%%%%%%%
In the real financial market  investors can held traded risky assets of maturity time $T$ and receive some particular derivatives such as contingent claims offering some pay-off at maturity time $T'>T>0$. It can happen that the assets related with contingent claims can not be traded since the trading is difficult or impossible for investor because of lack of liquidity or legal restrictions.
In this situation the investor would like maximize expected utility of total wealth and at the same time reduce the risk due to the uncertainty of pay-off of the contingent claim. In such situations the utility indifference pricing
become to be a main tool for option pricing.
\par To be more precise, let us suppose that our market consists on non-risky asset
$B_t= B_0\exp (rt)$, where $r$ is interest rate, and two risky assets
$$S_t= S_0 \,\mathcal E(X)_t, \,\,\, \tilde{S}_t= \tilde{S}_0 \,\mathcal E(\tilde{X})_t$$
where $X$ and $\tilde{X}$ are semi-martingales with jumps $\Delta X> -1,\,\Delta \tilde{X}>~-1$, and $\mathcal E$ is Dolean-Dade exponential. The investor can trade $S$
and at the same time he has a European type claim on $\tilde{S}$ given by $g(\tilde{S}_{T'})$ where $g$ is some real-valued Borel function. Let us denote by $\Pi$ the set of self-financing admissible strategies. Then, for utility function $U$ and initial capital $x$, the optimal expected utility related with $S$  will be
$$V_T(x) = \sup_{\phi\in \Pi} E [U(x+\int_0^T\phi_s\,dS_s)]$$
and if we add an option, then the optimal utility will be equal to
$$V_T(x,g) = \sup_{\phi\in \Pi} E [U(x+\int_0^T\phi_s\,dS_s\,+\,g(\tilde{S}_{T'})]$$
As known, the indifference price $p_T^b$ for buyer of the option $g(\tilde{S}_{T'})$
is a solution to the equation
$$V_T(x-p_T^b,g)=V_T(x)$$
and it is an amount of  money which the investor would be willing to pay today for the right to receive the claim and such that he is no worse off in expected utility terms then he would have been without the claim. The indifference price for the seller $p_T^s$ of the option is a solution to the equation
$$V_T(x+p_T^s,-g)=V_T(x)$$
and it is an amount of money which the seller of the option would be willing to receive in counterpart of the option in order to preserve his own optimal utility.
\par The optimal utility of assets containing the options highly depends on the level of information of the investor about  $\tilde{S}$. More precisely, the investor can  be non-informed, partially informed or perfectly informed agent and the level of information changes the class $\Pi$ mentioned in previous formulas. Namely, a non-informed agent can maximize his expected utility taking the strategies only from the set of self-financing admissible strategies with respect to the natural filtration
${\bf  F}$ of $X$. At the same time, a partially informed agent can built his optimal strategy using the set of  self-financing admissible strategies with respect to the progressively enlarged filtration $\tilde{\bf  F}$ with the process $\tilde{X}$. Finally, a perfectly informed agent can use the self-financing admissible strategies
with respect to initially enlarged filtration ${\bf G}$ with $\tilde{S}_{T'}$.
\par Utility maximisation and utility indifference pricing was considered in a number of books and papers, see for instance \cite{BJ}, \cite{BFG},\cite{Ca}, \cite{DGR}, \cite{HeH}, \cite{KrS}, \cite{MZ1}, \cite{MZ2}, \cite{MSch},\cite{RK}. Some explicit formulas for the indifference prices were obtained for Brownian motion models, where the incompleteness on the market comes from the non-traded asset (see \cite{HeH}, \cite{MZ1}, \cite{MZ2}). Close to our setting case, but for complete markets, was considered in \cite{ABSch} and one can find there nice explicit formulas for indifference price.
\par In this note we concentrate ourselves on non-complete market case, and we establish some explicit formulas for the indifference prices for semi-martingale models when the traded and non-traded asset are dependent. This dependence is modelled by including the non-traded asset into the structure of the traded asset as a factor influencing  its price dynamics. We will concentrate ourselves on the problem of utility maximisation and utility indifference pricing for perfectly informed agents.
Our aim is to obtain explicit and numerically tractable solutions for these questions, especially for exponential Levy models and diffusions. 
\par It should be noticed that the indifference price for partially informed agents and non-informed agents will be the same in considered case since the $\sigma$-algebras at time $T$ in all three cases coincide( to do calculus, one has to ensure that $g(\xi)$ is measurable!), and the last fact implies that the minimal equivalent martingale measure, when exist, will be the same in three cases, too. Contrarily to this, the optimal strategies will  depend on the used filtration. It should be noticed that in the case of exponential Levy models and HARA utilities the optimal strategy for initial enlargement, when it exists, is always progressively adapted. The same is true for the processes with independent increments. This fact can be explained by preservation of Levy property and "independent increments" property by minimal equivalent martingale measure when  such measure exists,
and explicit formulas for optimal strategies( cf.\cite{CV1},\cite{CV2}). 
\par From point of view of modelling our approach consists to introduce   semi-martingales depending on a random factor $\xi$. Namely, the considered risky asset $S$ will be of the form $S(\xi)=\mathcal{E}\left(X(\xi)\right)$ with the semi-martingale
$X(\xi)=\left(X_t(\xi)\right)_{t\geq 0}$ leaving  on  a canonical probability space  and depending on a supplementary random factor $\xi$. The random variable $\xi$  is given on Polish space  $(\Xi, \mathcal{H})$. We denote by $\alpha$  the law of this variable $\xi$. The details concerning such mathematical framework is given in section \ref{s1} and they are close to the approach in \cite{GVV}.
\par The section \ref{s2} is devoted to the general results about the maximisation of utility for semi-martingale models depending on a random factor.
As previously let us introduce the total utility  with the option $g(\xi)$:
\begin{equation*}
V(x,g)=\sup_{\varphi\in {\Pi(\bf G)}}E_{\mathbb{P}}\left[U\left(x+\int_{0}^{T}{\varphi_{s} dS_{s}(\xi)}+g(\xi)\right)\right]
\end{equation*}
Here $\Pi ({\bf G})$ is the set of all self-financing  and admissible trading strategies related with the initially enlarged filtration ${\bf{G}}=(\mathcal{G}_t)_{t\in\left[0,T\right]}$, where $\mathcal{G}_t=\bigcap_{s>t}\left(\mathcal{F}_s\otimes\sigma(\xi)\right)$. To solve the utility maximisation problem in the initially enlarged filtration  we make an  assumption about the absolute continuity  of the conditional laws $\alpha^t= \mathbb{P}(\xi\,|\, \mathcal F _t)$ of the random variable $\xi$ given $\mathcal{F}_t$  with respect  to $\alpha$, namely
$$\alpha^t <\!\!< \alpha$$
for $t\in ]0,T]$.
 Then we define the conditional laws $(P^u)_{u\in\Xi}$ of our semi-martingale $S(\xi)$ given $\{\xi=u \}$  and we reduce the initial  utility maximisation problem to the conditional utility maximisation problem on the asset prices filtration ${\bf F}$ ( see Proposition \ref{t1}).
Proposition \ref{t1} says that  to solve the utility maximisation problem on the enlarged filtration it is enough to solve the conditional utility maximisation problem  on the asset prices filtration ${\bf F}$ 
\begin{equation*}
V^u(x,g)=\sup_{\varphi\in {\Pi ^u({\bf F})}}E_{P^u}\left[U\left(x+\int_{0}^{T}{\varphi_{s}(u) dS_{s}(u)}+g(u)\right)\right]
\end{equation*}
and then integrate the solution with respect to  $\alpha$. To solve conditional utility maximisation problem we use dual approach. Let us denote by $f$ a convex conjugate of $U$. Under the assumption  about the existence of an equivalent $f$-divergence minimal measure for the conditional semi-martingale model, we give the expression for conditional maximal utility (cf. Proposition \ref{t2}). The main result of this section is
Theorem \ref{t3} which gives the final result for general utility maximisation
problem.
\par In section \ref{s3} we study HARA utilities. For HARA utilities we introduce corresponding information quantities and we give the expression for the maximal expected utility in terms of these quantities (cf. Theorem \ref{t4}). Finally, we introduce the information processes and we give the expression of  the maximal expected utility involving these information processes (see Propositions \ref{p1}, \ref{p2}, \ref{p3} and Theorem \ref{t5}).
\par In section \ref{s4} we give the formulas for indifference price of buyers and sellers of the option for HARA utilities. Then we discuss risk measure properties of the mentioned indifference prices. We show that $-p^b_T(g)$ and $p^s_T(g)$ are risk measures.
\par In the section \ref{s5} we study utility maximisation and utility
indifference pricing of exponential Levy models. It should be noticed that in Levy models case the information processes are deterministic processes containing the constants which are the solutions of relatively simple integral equations. It gives  us the possibility to calculate the indifference prices relatively easy.
\par The section \ref{s6} in devoted to the explicit calculus of information quantities for Geometric Brownian motion case and use identity in law technique.
\end{section}

%%%%%%%%%%%%%%%%%%%%%%%%%%%%%%%%%%%%%%%%%%%%%%%%%%%%%%%%%%%%%%%%%%%%%%%%%%%%%%
\begin{section}{Mathematical Framework} \label{s1}
%%%%%%%%%%%%%%%%%%%%%%%%%%%%%%%%%%%%%%%%%%%%%%%%%%%%%%%%%%%%%%%%%%%%%%%%%%%%%%
\par We consider a semi-martingale $X(\xi)=\left(X_t(\xi)\right)_{t\geq 0}$ of the law $P$, depending on a supplementary factor $\xi$ which can be a random process or a random variable. The semi-martingale $X(\xi)$ is given on a canonical probability space $(\Omega,\mathcal{F},P)$, equipped with the filtration ${\bf F}=\left(\mathcal{F}_t\right)_{t\geq 0}$ satisfying the usual conditions: $\mathcal{F}=\bigvee_{t\geq 0}\mathcal{F}_t$, $\mathcal{F}_t=\bigcap_{u>t}\sigma\{X_v(\xi), v\leq u\}$ and $\mathcal{F}_0=\{\emptyset,\Omega\}.$
\par We suppose that the law $P$ of $X(\xi)$ is uniquely defined by its semi-martingale characteristics $(B,C,\nu)$. We recall here  the notion of the characteristics   for the convenience of the readers.
Let $\mu$ be a jump measure of the process $X=X(\xi)$  and $l:\mathbb{R}\rightarrow \mathbb{R}$ be a truncation function: $l(x)=x$ in the neighbourhood of $0$ and $l$ has a compact support. Then one can write the semi-martingale $X$ as
\begin{equation*}
X=(X-X(l))+X(l),
\end{equation*}
where $X(l)$ is a 'big' jumps process,  defined as 
\begin{equation*}
X(l)_t=\sum_{s\leq t}\left(\Delta X_s-l(\Delta X_s)\right)
\end{equation*}
with $\Delta X_s=X_s-X_{s-}.$ The process $\tilde{X}=\left(X-X(l)\right)$ is a special semi-martingale with the bounded jumps and allows the representation 
\begin{equation*}
\tilde{X}_t=X_0+X_t^c+\int_0^t\int_{\mathbb{R}} l(x)\left(\mu(ds,dx)-\nu(ds,dx)\right)+B_t(l),
\end{equation*}
where $X^c$ is the continuous local martingale part of $X$, $\nu$ is the $(P,{\bf F})$ compensator of $\mu$, $B=B(l)$ is the unique $(P,{\bf F})$-predictable locally integrable process such that the process $\tilde{X}-B(l)$ is a $(P,{\bf F})$-local martingale. Let $C$ be a continuous process such that the process $(X^c)^2-C$ is a $(P,{\bf F})$-local martingale. We have defined the triplet of predictable characteristics  of the $(P,{\bf F})$-semi-martingale $X=X(\xi)$ as $T^{{\bf F}}=(B, C,\nu)$ (see  also \cite{JSh}).
\par We suppose that the supplementary random factor $\xi$ is given on the  probability space $(\Xi, \mathcal{H},\alpha)$ with  $\alpha$ being the law of $\xi$.
\par We assume that our market contains a single traded risky asset with the price process $S=S(\xi)$ and without any loss of generality we will assume that the riskless interest rate is $0$, and then the riskless bond process is identically equal to $1$. Our risky asset  $S=S(\xi)$ which we consider will be simply of the form
\begin{equation}
S(\xi)=\mathcal{E}\left(X(\xi)\right),
\label{1}
\end{equation}
where $\mathcal{E}(\cdot)$ is a stochastic exponential,
\begin{equation*}
\mathcal{E}(X)_t=\exp\big\{X_t-\frac{1}{2}<X^c>_t\big\}\prod_{0\leq s\leq t} \exp\{-\Delta X_s\}(1+\Delta X_s).
\end{equation*}
We prefer the representation \eqref{1} of the risky asset more than the representation with the usual exponent  by the simple reason that  if  the process $X$ is a local $(P,{\bf F})$-martingale then the process $S$ inherits this property, i.e. is a local $(P,{\bf F})$-martingale. To ensure that $S_t> 0$ for all $t\geq 0$ we assume that $\Delta X_t> -1$.
\par The process $X(\xi)$ can be defined in a different way.
One of the possibilities is to give, when they exist, a family of the regular conditional laws of $X(\xi)$ given $\xi=u$, denoted by $(P^u)_{u\in\Xi}$.
Such family of conditional laws should verify: for all $t\geq 0$,  for all $A\in\mathcal{F}$
\begin{equation}
P(A)=\int_{\Xi}P^u(A)\,d\alpha(u). 
\label{3a}
\end{equation}
\par Now, on the product space $(\Omega\times\Xi,\mathcal{F}\otimes\mathcal{H})$ we can also define a probability $\mathbb{P}$ as it follows: for all $ A\in\mathcal{F}$ and $B\in\mathcal{H}$
\begin{equation}
\mathbb{P}(A\times B)=\int_{B} P^u(A)d\alpha(u),
\label{3}
\end{equation}
such that $\mathbb{P}(A\times \Xi)=P(A)$ and $\mathbb{P}(\Omega\times B)=\alpha(B).$ In such situation for all $A\in \mathcal F$
$$P^u(A) = \mathbb{P}(A\,|\,\xi=u)$$
\par Now we define the initially enlarged filtration
${\bf G}=(\mathcal{G}_t)_{t\geq 0}$ by 
\begin{equation}
\mathcal{G}_t=\bigcap_{s>t}\left(\mathcal{F}_s\otimes\sigma(\xi)\right).
\label{6}
\end{equation}
Let $t\in\mathbb{R}_{+}$ and $\alpha^t$ be a regular conditional distribution of the random variable $\xi$ given the information $\mathcal{F}_t$, i.e. 
\begin{equation*}
\alpha^t(\omega, du)=\mathbb{P}(\xi\in du|\mathcal{F}_t)(\omega).
\end{equation*}
We make the following assumption
\asm \label{a1}The regular conditional distribution of random variable $\xi$ is absolutely continuous with respect to its law, i.e.
$$\alpha^t\ll\alpha,\ \  \forall t\in\left]0,T\right].$$

%%%%%%%%%%%%%%%%%%%%%%%%%%%%%%%%%%%%%%%%%%%%%%%%%%%%%%%%%%%%%%%%%%%%%%%%%%%%%%
\lem (see\cite{J1})\label{l1} Under Assumption \ref{a1}  there exists a positive $\mathcal{O}(\mathbb{G})$ measurable function $(\omega, t, u)\rightarrow p^u_t(\omega)$ such that 
 \begin{enumerate}
 \item For each $u\in\text{supp}(\alpha)$, $p^u$ is  $(P,{\bf F})$-martingale.
 \item For each $t\in\left[0,T\right]$, the measure $p^u_t\alpha(du)$ is a version of the regular conditional distribution $\alpha^t(du)$ so that $P_t\times\alpha$-a.s.
 \begin{equation}
 \frac{d\alpha^t}{d\alpha}(u)=p_t^u.
 \label{38}
 \end{equation}
 \end{enumerate}
\normalfont
%%%%%%%%%%%%%%%%%%%%%%%%%%%%%%%%%%%%%%%%%%%%%%%%%%%%%%%%%%%%%%%%%%%%%%%%%%%%%%
To avoid unnecessary complications, we introduce also
\asm \label{a11} For each $u\in\Xi$ the probability $P^u$ is locally absolutely continuous with respect to $P$, i.e
$$P^u\stackrel{loc}{\ll} P.$$
\rm

The Assumptions \ref{a1}, \ref{a11} and Lemma 1 imply that for all $t\in\left[0,T\right]$ and  $P_t\times\alpha$-a.s.
\begin{equation}
\frac{dP^u|_{\mathcal{F}_t}}{dP|_{\mathcal{F}_t}}=p_t^u
\label{2}
\end{equation}

The process $X$ is also $(P^u,{\bf F})$-semi-martingale, $u\in\Xi.$ If we know the density  $p^u$, then using Ito formula we can write the semi-martingale decomposition of it and restore the $(P^u,{\bf F})$-characteristic triplet $T^{{\bf F}}(u)=(B^u,C^u,\nu^u)$. This triplet is related to the triplet $T^{{\bf F}}=(B,C,\nu)$ as follows 
\begin{eqnarray}\label{5}
B^u&=&B+\int_0^{\cdot}\beta_s^u\,dC_s+\int_0^{\cdot}\int_{\mathbb R}l(x)\,(Y_s^u(x)-1) \nu (ds, dx),\nonumber\\
C^u&=&C,\nonumber\\
\nu^u&=&Y^u\cdot \nu,
\end{eqnarray}
with certain $(P^u,{\bf F})$-predictable process $\beta^u=(\beta^u_t)_{t\in\left[0,T\right]}$ and $Y^u=(Y^u_t)_{t\in\left[0,T\right]}$ such that $P-a.s$ for all ${t\in\left[0,T\right]}$
\begin{equation*}
\int _0^t (\beta^u_s)^2\,dC_s +\int _0^t\int_{\mathbb{R}}|\,l(x)\,(Y_s^u(x)-1)|\, \nu(ds,dx)<\infty.
\end{equation*}
For the details about the integration with respect to the random measures and its compensators , stochastic integration  with respect to a local martingales and  Riemann-Stieltjes integral see \cite{JSh}.
\par Since  the density process $p^u$ is a $(P,{\bf F})$-martingale,  we define the stochastic logarithm $m^u$ of $p^u$ by:
\begin{equation*}
dm^u_t=\frac{dp^u_t}{p^u_{t-}.}
\end{equation*}
Then $m^u$ is a $(P,{\bf F})$-local martingale and $p^u$ is a stochastic exponential of $m^u$
\begin{equation*}
p^u=\mathcal{E}(m^u).
\end{equation*}
By the predictable representation property we have that the local martingale $m^u$ has the following semi-martingale representation 
\begin{equation*}
m^u=\int_0^{\cdot}\beta _s^u\,dX_s^c+\int_0^{\cdot}\int_{\mathbb R}\left(Y_s^u-1+\frac{\hat{Y}_s^u-\hat{1}}{1-\hat{1}}\right) (\mu-\nu)(ds, dx),
\end{equation*}
where the process $\beta^u$ and $Y^u$ are the same as in \eqref{5} and the processes $\hat{Y}^u$ and $\hat{1}$  are related to the compensator $\nu$, namely 
\begin{equation*}
\hat{1}_t(\omega)=\nu(\omega,\{t\}\times\mathbb{R}_0)
\end{equation*}
and 
\begin{equation*}
\hat{Y}^u_t(\omega)=\int_{\mathbb{R}_0}{Y}^u_t(\omega,x)\nu(\omega,\{t\},dx).
\end{equation*}
For more information  see again \cite{JSh}. 
\end{section}
%%%%%%%%%%%%%%%%%%%%%%%%%%%%%%%%%%%%%%%%%%%%%%%%%%%%%%%%%%%%%%%%%%%%%%%%%%%%%%
\begin{section}{Utility maximisation problem }\label{s2}%%%%%%%%%%%%%%%%%%%%%%
%%%%%%%%%%%%%%%%%%%%%%%%%%%%%%%%%%%%%%%%%%%%%%%%%%%%%%%%%%%%%%%%%%%%%%%%%%%%%%

In this section we introduce the sets of the self-financing admissible trading strategies and the sets of the equivalent martingale measures for the initially enlarged filtration and we establish the connection between them and the analogous sets on the $(\Omega, \mathcal{F}, {\bf F}, P^u)$ filtered space. Then we show that  the solution of the utility maximisation problem in the enlarged filtration can be reduced to the solution of the conditional utility maximisation problem (cf. Proposition \ref{t1}) which in turn, we solve using the dual approach (cf. Proposition \ref{t2}). The final result on utility maximisation is given in Theorem \ref{t3} at the end of this section.
%%%%%%%%%%%%%%%%%%%%%%%%%%%%%%%%%%%%%%%%%%%%%%%%%%%%%%%%%%%%%%%%%%%%%%%%%%%%%%  
\subsection{Utility maximisation problem in enlarged filtration}\label{s3}
%%%%%%%%%%%%%%%%%%%%%%%%%%%%%%%%%%%%%%%%%%%%%%%%%%%%%%%%%%%%%%%%%%%%%%%%%%%%%%
We consider  a utility function $U:\mathbb{R}\rightarrow\mathbb{R}\bigcup\{-\infty\}$, which is assumed to be strictly increasing, strictly concave, continuously differentiable in $dom(U)=\{x\in\mathbb{R}|U(x)>-\infty\}$ and is supposed to satisfy the Inada conditions 
\begin{eqnarray*}
U^{'}(\infty)=\lim_{x\rightarrow +\infty}U^{'}(x)=0,\\
U^{'}(\overline{x})=\lim_{x\downarrow \underline{x}}U^{'}(x)=\infty ,
\end{eqnarray*}
where $\underline{x}=\inf\{x\in\mathbb{R}|U(x)>-\infty\}.$ We require that the utility function is the increasing function of the wealth because with the growth of wealth the investor's usefulness also grows. The concavity of the function reflects  a phenomenon of  risk-aversion for the investor. 
\par Suppose that the investor carries out the trading on the finite time interval $\left[0,T\right]$ and holds a European type option with the pay-off function $G_T=g(\xi)$ in his portfolio,  where $g$ is an $\mathcal{H}$-measurable function. We define by $\Pi(\bf{G})$ the set of admissible and self-financing strategies $\varphi(\xi)$, such that $\varphi(\xi)$ is $\bf{G}$-predictable and $S(\xi)$-integrable on $[0,T]$ $P-a.s.$,  with the integrals bounded from below. To describe this set we recall the known result about ${\bf G}$-predictable processes, denoted by $\mathcal P ({\bf G})$.
\lem \label{l2}
%%%%%%%%%%%%%%%%%%%%%%%%%%%%%%%%%%%%%%%%%%%%%%%%%%%%%%%%%%%%%%%%%%%
 $\left(\text{cf. }\left[4\right]\right)$
A random process $\varphi(\xi)$ is ${\bf{G}}$-predictable if and only if 
the application $(t,\omega,\xi)\rightarrow \varphi_t(\xi)$
 is a $\mathcal{P}\left({\bf F}\right)\otimes\mathcal{H}$-measurable random process where $\mathcal P({\bf F})$ is the set of ${\bf F}$-predictable processes.
\normalfont 
\par Thus,  the set  of the admissible and self-financing strategies  $\Pi(\bf{G})$ on $(P, \bf{G})$ is of the form
\begin{multline*}
\Pi({\bf{G}})=\bigcup_{c>0}\big\{\varphi(\xi)\in \mathcal P({\bf F})\otimes \mathcal H\,|\, \int _0^t\varphi_s(\xi) dS_s(\xi)\geq-c,\, \forall t\in\left[0,T\right](\text{\,$\mathbb{P}$-a.s.)}
 \big\}
% \label{46}
\end{multline*}
\par The classical utility maximisation problem consists to find the optimal investment portfolio over set of all self-financing and admissible strategies in order to maximise the given expected utility, namely
\begin{equation}
\label{9}
V(x,g)=\sup_{\varphi \in  \Pi(\bf{G}) }
E_{\mathbb{P}}\left[U\left(x+\int_{0}^{T}{\varphi_{s}(\xi) dS_{s}{(\xi)}}+g(\xi)\right)\right],
\end{equation}
where $U$ is the given utility function and $x$ is the initial endowment. 
\par We define also the set  $\Pi^u({\bf F})$ of the admissible and self-financing strategies related with the filtration ${\bf F}$:
\begin{equation*}
 \Pi^u({\bf F})=\bigcup_{c>0}\big\{\varphi\in S_u(\mathcal{P}({\bf F})\otimes \mathcal H ) \,|\, \int _0^t\varphi_s(\xi)dS_s(\xi )\geq-c,\, \forall t\in\left[0,T\right](\text{\,$\mathbb{P}$-a.s.)}\big\}
\end{equation*}
where $S_u(\mathcal{P}({\bf F})\otimes \mathcal H )$ is a section of $\mathcal{P}({\bf F})\otimes \mathcal H$ in $u$.
For any $u\in\Xi$ we denote
\begin{equation}
\label{48}
V^{u}(x,g)=\sup_{\varphi \in  \Pi^u({\bf F}) }
E_{P^{u}}\left[U\left(x+\int_{0}^{T}{\varphi_{s}(u) dS_{s}(u)}+g(u)\right)\right]
\end{equation}
\par The next result  establishes  that the value of the maximal utility in enlarged filtration  $\bf{G}$  can be obtained  from the solutions of the conditional utility maximisation problem. 
%%%%%%%%%%%%%%%%%%%%%%%%%%%%%%%%%%%%%%%%%%%%%%%%%%%%%%%%%%%%%%%%%%%%%%%%%%%%% 
 \prop \label{t1}
%%%%%%%%%%%%%%%%%%%%%%%%%%%%%%%%%%%%%%%%%%%%%%%%%%%%%%%%%%%%%%%%%%%%%%%%%%% 
Let us suppose that Assumptions \ref{a1} and \ref{a11} hold. Then we can reduce classical utility maximisation problem to the corresponding conditional utility maximisation problem in the sense that 
 \begin{equation}
 V(x,g)=\int_{\Xi}V^{u}(x,g)d\alpha(u).
 \label{56}
 \end{equation}

\normalfont 
%%%%%%%%%%%%%%%%%%%%%%%%%%%%%%%%%%%%%%%%%%%%%%%%%%%%%%%%%%%%%%%%%%%
To prove this proposition we prove first one lemma.
%%%%%%%%%%%%%%%%%%%%%%%%%%%%%%%%%%%%%%%%%%%%%%%%%%%%%%%%%%%%%%%%%%%

%%%%%%%%%%%%%%%%%%%%%%%%%%%%%%%%%%%%%%%%%%%%%%%%%%%%%%%%%%%%%%%%%%%
\lem \label{l3}
%%%%%%%%%%%%%%%%%%%%%%%%%%%%%%%%%%%%%%%%%%%%%%%%%%%%%%%%%%%%%%%%%%%
Let $\varphi(\xi)\in\Pi(\bf{G})$. Then for $t\in [0,T]$ and $u\in \Xi$
\begin{equation}
\mathcal{L}_{\mathbb{P}}\left((\int_0^t\varphi_s(\xi)dS_s(\xi), \xi)\Big{|}\xi=u\right)=
\mathcal{L}_{P^u}\left((\int_0^t\varphi_s(u)dS_s(u), u)\right).
\label{49}
\end{equation}
As consequence, we get that
\begin{equation*}
E_{\mathbb{P}}\left[U\left(x+\int_0^t\varphi_s(\xi)dS_s(\xi)+ g(\xi)\right)\Big{|}\xi=u\right]=
E_{P^u}\left[U\left(x+\int_0^t\varphi_s(u)dS_s(u)+g(u)\right)\right].
%\label{50}
\end{equation*}

\pf\normalfont
It is known that $\Pi(\bf{G})$ can be generated by the simple  functions of the type $\varphi(\xi)=1_{A}(\xi)\varphi_{t_1}1_{\left]t_1,t_2\right]} $, where $t_1,t_2\in\mathbb{R}_{+},\ \ t_1\leq t_2,\ \ A\in\mathcal{H}$ and $\varphi_{t_1}$ is $\mathcal{F}_{t_1}$-measurable random variable. For such $\varphi(\xi)$ we have:
\begin{equation*}
\int_0^T\varphi_s(\xi)dS_s(\xi)=1_{A}(\xi)\varphi_{t_1}(S_{t_2}(\xi)-S_{t_1}(\xi)).
\end{equation*}
Since the filtration ${\bf F}$ is natural, $\varphi_{t_1}=F(X_v(\xi),0\leq v \leq T)$ where $F$ is a measurable functional. Since 
\begin{equation}
\mathcal{L}_{\mathbb{P}}\left((\xi, X(\xi))|\xi=u\right)=\mathcal{L}_{P^u}\left(u, X\right),
\label{53}
\end{equation}
the same identity in law  is true  for measurable functional of $(\xi,X(\xi))$ such as $(S(\xi), \varphi_{t_1},1_{A}(\xi))$ and it gives \eqref{49} for a special type of $\varphi(\xi)$.
\par For general $\varphi(\xi)\in\Pi({\bf{G}})$ there exists a sequence of linear combination of simple functions, $(\varphi^{n}(\xi))_{n\in\mathbb{N}_{+}}$ such that for  $s\in\left[0,T\right]$ and $P\times\alpha$-a.s.
$$\varphi^n_s(\xi)\rightarrow \varphi_s(\xi),$$ 
and $|\varphi^n_s(\xi)|\leq|\varphi_s(\xi)|$. Since $\varphi(\xi)$ is locally bounded predictable function, then, according to Theorem I.4.31 in \cite{JSh}, we have the convergence in $P$-law: 
\begin{equation}
\int_0^T\varphi_s^{n}(\xi)dS_s(\xi)\underset{n\rightarrow \infty}{\longrightarrow}
\int_0^T\varphi_s(\xi)dS_s(\xi).
\label{51}
\end{equation} 
For the same reason and since  for  $s\in\left[0,T\right]$ ($P^u\times\alpha$-a.s)
$$\varphi_s^n(u)\longrightarrow \varphi_s(u),$$ 
we have the convergence in $P^u$-law:
\begin{equation}
\int_0^T\varphi_s^{n}(u)dS_s(u)\underset{n\rightarrow \infty}{\longrightarrow}
\int_0^T\varphi_s(u)dS_s(u).
\label{52}
\end{equation} 
From \eqref{53},  \eqref{51} and \eqref{52} we obtain \eqref{49}.
If we denote $\Phi(v,r)=U(x+v+g(r))$ then it is a $\mathcal{B}(\mathbb{R}^2)$-measurable function of $(v,r)$ for all $x\in\mathbb{R}_{+}$, and it gives the second claim. Then lemma is proved. $\Box$

\pff \normalfont
If in Lemma \ref{l3} we take  regular versions of stochastic integrals and conditional expectations ( cf. \cite{StY}),
then we have: 
\begin{eqnarray*}
&&E_{\mathbb{P}}\left[U\left(x+\int_0^T\varphi_s(\xi)dS_s(\xi)+g\left(\xi\right)\right)\right]=\\
&=&\int_{\Xi}E_{P^u}\left[U\left(x+\int_0^T\varphi_s(u)dS_s(u)+g(u)\right)\right]d\alpha(u)\\
&\leq& \int_{\Xi}sup_{\varphi \in\Pi^u({\bf F})}E_{P^u}\left[U\left(x+\int_0^T\varphi_s(u)dS_s(u)+g(u)\right)\right]d\alpha(u),
\end{eqnarray*}
and hence,
\begin{equation}
V(x,g)\leq\int_{\Xi} V^u(x,g)d\alpha(u).
\label{55}
\end{equation}
\par For each $\epsilon>0$ there exists $\varphi^{(\epsilon)}\in\Pi^u(\bf{F})$ such that 
\begin{equation*}
sup_{\varphi \in\Pi ^u(\bf{F})}E_{P^u}\left[U\left(x+\int_0^T\varphi_s(u)dS_s(u)+g(u)\right)\right]\leq 
\end{equation*}
$$E_{P^u}\left[U\left(x+\int_0^T\varphi^{(\epsilon)}_s(u)dS_s(u)+g(u)\right)\right]+\epsilon$$

Integration with respect to $\alpha$ gives:
\begin{eqnarray*}
&&\int_{\Xi}sup_{\varphi \in\Pi ^u(\bf{F})}E_{P^u}\left[U\left(x+\int_0^T\varphi_s(u)dS_s(u)+g(u)\right)\right]d\alpha(u)\leq\nonumber\\
&\leq&\int_{\Xi}E_{P^u}\left[U\left(x+\int_0^T\varphi^{(\epsilon)}_s(u)dS_s(u)+g(u)\right)\right]d\alpha(u)+\epsilon\nonumber\\
&=&E_{\mathbb{P}}\left[U\left(x+\int_0^T\varphi^{(\epsilon)}(\xi)_sdS_s(\xi)+g(\xi)\right)\right]+\epsilon\nonumber\\
&\leq&V(x,g)+\epsilon
\label{54}
\end{eqnarray*}
Combining the two previous inequalities  we have \eqref{56}.
$\Box$
%%%%%%%%%%%%%%%%%%%%%%%%%%%%%%%%%%%%%%%%%%%%%%%%%%%%%%%%%%%%%%%%%%%%%%%%%%%%
\subsection{The solution to conditional utility maximisation problem}
%%%%%%%%%%%%%%%%%%%%%%%%%%%%%%%%%%%%%%%%%%%%%%%%%%%%%%%%%%%%%%%%%%%%%%%%%%%%

\par To solve the conditional utility maximisation problem $V^u(x,g)$ we use the dual approach. For that we consider the equivalent martingale measures in the enlarged  filtration ${\bf G}$ and then we provide the link between them and the equivalent martingale measures related to $(P^u,{\bf F}).$
\par Let $\mathcal{M}(\bf{G})$ be a set of $\mathbb{P}$-equivalent martingale measures on product space $(\Omega\times\Xi,\mathcal{F}\otimes\mathcal{H})$ defined as
\begin{multline*}
\mathcal{M}({\bf{G}})=\big\{\mathbb{Q}: \ {\mathbb{Q}}\stackrel{loc}{\sim} \mathbb{P}\ \text{and}\ \text{such that } S(\xi) \text{ is an $\left({\mathbb{Q}},\bf{G}\right)$-martingale}\big\}.
\end{multline*}
\par Let $T$ be a finite time horizon. Then the restrictions of the measures $\mathbb{Q}$ on the $\sigma$-algebra $\mathcal{G}_T$ can be given by  density process $Z(\xi)$:
\begin{equation}
\frac{d{\mathbb{Q}}|_{\mathcal{G}_T}}{d{\mathbb{P}}|_{\mathcal{G}_T}}=Z_T(\xi)
\label{14}
\end{equation}
The density process $Z(\xi)=(Z_t(\xi))_{t\in\left[0,T\right]}$ is a uniformly integrable positive $(\mathbb{P},\bf{G})$-martingale  with $E_{\mathbb{P}}\left[{Z}_T(\xi)\right]=1.$
We recall the following known result about ${\bf G}$-martingales.  Let us fix $u\in supp(\alpha)$ and let the process $Z(u)$ be obtained from the process $Z(\xi)$ by replacing of $\xi$ by $u$.
\lem  $\left(\text{cf. }\left[4\right]\right)$  Under Assumptions \ref{a1} and \ref{a11} there exists a version of density process $Z(\xi)$ such that the following two statements are equivalent:

$(i)$ The process $Z=Z(\xi)$ is a $(\mathbb{P},\bf{G})$-martingale

$(ii)$ The process ${Z}(u)=(Z_t(u))_{t\in\left[0,T\right] }$ is a $(P^u,{\bf F})$-martingale, for all $u\in\text{supp}(\alpha)$.

\normalfont
\par As it was mentioned, $Z(u)$ is a positive  $(P^u,{\bf F})$-martingale. However $Z(u)$ is not a density process because of the fact that $$E_P\left[Z_T(u)\right]=Z_0(u)$$
with $Z_0(u)$ which is not necessarily equal to $1$.  
But the modified density process process $\tilde{Z}(u)=\frac{Z(u)}{Z_0(u)}$ describes the equivalent martingale measures $Q^u$
such that 
\begin{equation}
\label{eq37}
\frac{d{Q}^u|_{\mathcal{F}_t}}{dP^u|_{\mathcal{F}_t}}=\tilde{Z}_t(u).
\end{equation}
We denote by $\mathcal{M}^u({\bf F})$ the set of such measures, namely
\begin{multline}
\label{eq39}
\mathcal{M}^u({\bf F})=\big\{{Q}^u: \ {Q}^u\stackrel{loc}{\sim}  P^u,\ \ S \text{ is an $\left({Q}^u,{\bf F}\right)$-martingale}\big\}
\end{multline}
\par Let us denote by $f$ the convex conjugate of $U$ obtained  by Frenchel-Legendre transform of $U$:  
\begin{equation*}
f(y)=\underset{x>0}{\sup}\left(U(x)-yx\right).
\end{equation*}
Let us denote by $I(y)=-f^{'}(y),\ \ y\in\mathbb{R}_{+}$, then
\begin{equation}
f(y)=U(I(y))-yI(y),
\label{eq5}
\end{equation}

\par Now we consider the dual problem of finding 
\begin{equation*}
\inf_{Q^u\in\mathcal{M}^u({\bf F})}E_{P^u}\left[f\left(\frac{dQ^u_T}{dP_T}\right)\right].
\end{equation*}
If minimum is reached on the set $\mathcal{M}^u({\bf F})$, then the corresponding measure $Q^{u, *}$ is called $f$-divergence minimal martingale measure.
\par Let also $u\in\Xi$ to be fixed, and the set $\mathcal{K}^u$ be defined as follows:
\begin{multline*}
\mathcal{K}^u=\Big\{Q^u\in\mathcal{M}^u({\bf F}):\,E_{P^u}\left|f\left(\lambda\frac{ dQ^u_T}{dP_T}\right)\right| < \infty,\ E_{Q^u}\left|f^{'}\left(\lambda \frac{dQ^u_T}{dP_T}\right)\right| < \infty ,\forall\lambda>0\Big\}.
\end{multline*}
We introduce  two additional Assumptions.
\asm \label{a2} For each $u\in \Xi$, there exists $f$-divergence minimal equivalent martingale measure $Q^{u, *}$, it belongs to the set $\mathcal{K}^{u}$ and verify scaling property: for each $\lambda >0$
$$\inf_{Q^u\in\mathcal{M}^u({\bf F})}E_{P^u}\left[f\left(\lambda\,\frac{dQ^u_T}{dP_T}\right)\right]= E_{P^u}\left[f\left(\lambda\,\frac{dQ^{u,*}_T}{dP_T}\right)\right]. $$

\rem For HARA utilities the scaling property is automatically verified and the definition of the set $\mathcal{K}^u$ can be simplified:
\begin{equation*}
\mathcal{K}^u=\Big\{Q^u\in\mathcal{M}^u({\bf F}):\,E_{P^u} |f\left(\frac{ dQ^u_T}{dP_T}\right)| < \infty\Big\}.
\end{equation*}
\normalfont
The next assumption is related with the properties of the density $\tilde{Z}^*_T(u)$ of $f$-divergence minimal equivalent martingale measure  $Q^{u, *}_T$  with respect to $P_T^u$.
\asm \label{a3}  There exists $\mathcal H$- measurable function $\lambda _{g}$, which verifies:
$$\int_{\Xi}E_{P^u} |f(\lambda _{g}(u)\,\tilde{Z}_T^*(u)| d\alpha (u)<\infty,\,\,\int_{\Xi}E_{Q^u}  |f'(\lambda _{g}(u)\,\tilde{Z}_T^*(u)| d\alpha (u)<\infty$$
and such that for each $u\in \Sigma$ and $x> \underline{x}$
\begin{equation}
E_{P^u}\left[\tilde{Z}^{*}_T(u)\,I(\,\lambda _{g}(u)\tilde{Z}_T^{*}(u)\,)\right]=x+g(u).
\label{62}
\end{equation}
\normalfont

\rem For HARA utilities the integrability conditions of the Assumption \ref{a3} is  reduced to the first one. 
\normalfont
%\rem 
%Let us show that there exists $\mathcal H$-measurable modification of $\lambda$ %verifying \eqref{62}. For that we take an increasing sequence of peace-wise %constant $\mathcal H$-measurable functions $(g^{(n)})_{n\in \mathbb N}$ such that %$g^{(n)}(u)\rightarrow g(u)$ as $n\rightarrow \infty$. Let for $u\in \Xi$, %$g^{(n)}(u) = \sum _{i=1}^n \alpha _i  1_{U_i}(u)$ with $\alpha _i \geq 0$ and %$U_i \in \mathcal H$. Since the solution of \eqref{62} exists for all $x> %\underline{x}$, we deduce that there exists
%$\lambda ^{(n)}= \sum _{i=1}^n\lambda _i 1_{U_i}$ the solution of \eqref{62}
%with $g^{(n)}$. Since $I= -f'$ and $f$ is strictly convex function, $I$ is %strictly decreasing function and this implies that  the sequence $(\lambda %^{(n)})_{n\in \mathbb N}$ is positive and decreasing. So, for a.e. $u$ it %converges to a limit $\tilde{\lambda}$ which is $\mathcal H$-measurable. Using %Lebesgue monotone convergence theorem we prove that \eqref{62} holds for this %modification $\tilde{\lambda}$.
%\normalfont

%%%%%%%%%%%%%%%%%%%%%%%%%%%%%%%%%%%%%%%%%%%%%%%%%%%%%%%%%%%%%%%%%%%%%%%%%%%%%%%
\prop \label{t2}
%%%%%%%%%%%%%%%%%%%%%%%%%%%%%%%%%%%%%%%%%%%%%%%%%%%%%%%%%%%%%%%%%%%%%%%%%%%%%%
Let the Assumptions \ref{a2} and \ref{a3} hold.
Then there exists an optimal strategy $\varphi\in\Pi^u({\bf F})$ such that 
\begin{equation*}
V^u(x,g)=E_{P^u}\left[U\left(x+\int_0^T\varphi^{*}_s(u)dS_s(u)+g(u)\right)\right]
\end{equation*}
Moreover,  we have 
\begin{equation}
V^{u}(x,g)=E_{P^{u}}\left[U\left(I(\,\lambda(u)\tilde{Z}_T^{*}(u)\,)\right)\right]
\label{61}
\end{equation}
\pf\normalfont
%%%%%%%%%%%%%%%%%%%%%%%%%%%%%%%%%%%%%%%%%%%%%%%%%%%%%%%%%%%%%%%%%%%%%%%%%%%%%%
For any martingale measure $\mathbb Q_T$ equivalent to $\mathbb P_T$, and $Z_T(\xi)$ its Radon-Nikodym derivative which is $\mathcal F\otimes \mathcal H$-measurable, we write:
$$E_{\mathbb P }\,f(Z_T(\xi)) = \int_{\Xi}E_{P^u}\,f(Z_T(u))\, d\alpha (u) = $$
$$\int_{\Xi}E_{P^u}\,f(Z_0(u)\tilde{Z}_T(u))\, d\alpha (u)$$
where $\tilde{Z}_T(u))= Z_T(u) /Z_0(u)$.
Now, we consider conditional $f$-divergence minimisation problem, i.e. find
$\inf E_{P^u}\,f(\tilde{Z}_T(u))$ over all  martingale measures $Q^u$ equivalent to $P^u$ with $\tilde{Z}_T(u) = \frac{dQ^u_T}{dP^u_T}$, under initial capital equal to  $x+g(u)$. Let $Q^{u,*}_T$ be f-minimal equivalent martingale measure and 
 $\tilde{Z}^*_T(u) = \frac{dQ^{u,*}_T}{dP^u_T}$. According to Assumption \ref{a3}, there exists  
$\lambda _g(u)$ such that 
$$E_{Q^{u,*}}(\,I(\lambda _g(u)\,\tilde{Z}_T^*(u))= x + g(u)$$
One can show that $\lambda _g$ is unique $\alpha-a.s.$.
\par Since standard $f$-divergences verify scaling property, we get for any $\mathbb Q _T$:
$$E_{\mathbb P }\,f(Z_T(\xi)) \geq \int_{\Xi}E_{P^u}\,f(Z_0(u)\,\tilde{Z}_T^*(u))\, d\alpha (u) $$
But $Z_0(u)$ is entirely defined by the restriction on conditional initial capital, so the minimum over all equivalent martingale measures $\mathbb Q_T$ with this restriction is $\int_{\Xi}E_{P^u}\,f(\lambda _g(u)\tilde{Z}_T^*(u))\, d\alpha (u) $.

Then we can use the result of \cite{GR} and write ($\mathbb P$ - a.s.):
$$I(\lambda _g(\xi)\, \tilde{Z}^*_T(\xi)) = x +g(\xi) +\int_0^T \varphi_s^*(\xi)\,dS(\xi)$$
 where $\varphi^*\in \mathcal P({\bf G})$, it is self-financing and admissible, and such that $\int_0^{\cdot} \varphi_s^*(\xi)\,dS(\xi)$ is $\mathbb Q^*$-martingale.
The previous expression conditioned in $\xi=u$ gives ($P^u$-a.s.):
$$I(\lambda _g(u)\, \tilde{Z}^*_T(u)) = x +g(u) +\int_0^T \varphi_s^*(u)\,dS(u)$$
We see that $\varphi^*(u)\in \Pi^u({\bf F})$, it is self-financing, admissible and such that
$\int_0^{\cdot} \varphi_s^*(u)\,dS(u)$ is $ Q^{u, *}$-martingale.
\par Now we show that $\varphi ^*(u)$ is optimal strategy. Let us put
$x+g(u)= \tilde{x}$. Then, since \eqref{eq5} and $I(y) = -f'(y)$,
$$U(\tilde{x} + \int_0^T \varphi_s^*(u)\,dS(u))
 = f(\lambda _g(u) \tilde{Z}^*_T(u)) + \tilde{Z}^*_T(u)\,f'(\lambda _g(u)\,\tilde{Z}^*_T(u))$$
We show easily that the left-hand side of the previous equality is integrable:\\\\
$E_{P^u}|\,U(\tilde{x} + \int_0^T \varphi_s^*(u)\,dS(u))\,| \leq $
$$E_{P^u}| f(\lambda _g(u) \tilde{Z}^*_T(u))| + \,E_{P^u}|\tilde{Z}^*_T(u)\,f'(\lambda _g(u)\,\tilde{Z}^*_T(u))| < \infty$$
We write for any $\varphi\in \Pi ^u({\bf F})$ using the definition of  Fenchel-Legendre transform: 
$$U(\tilde{x} + \int_0^T \varphi_s(u)\,dS(u))\leq 
\left[\tilde{x} + 
\int_0^T \varphi_s(u)\,dS(u)\right]\lambda _g(u) \tilde{Z}_T^*(u)$$
$$ + f(\lambda _g(u) \tilde{Z}^*_T(u))
\leq \left[\tilde{x} + \int_0^T \varphi_s(u)\,dS(u)\right]\lambda _g(u) \tilde{Z}^*_T(u) +
U(I(\lambda _g(u)\, \tilde{Z}^*_T(u)) -$$
$$ \lambda _g(u)\, \tilde{Z}^*_T(u)\, I(\lambda _g(u)\, \tilde{Z}^*_T(u))$$
We take an expectation with respect to $P^u$, then we use the fact that
$ \int_0^{\cdot} \varphi_s(u)\,dS(u)$ is a super-martingale started from zero and 
that $ \int_0^{\cdot} \varphi ^*_s(u)\,dS(u)$ is a martingale with respect to $Q^{u,*}$.
Finally we get that
$$E_{P^u}[\,U(\tilde{x} + \int_0^T \varphi_s(u)\,dS(u))]\leq E_{P^u}[\,U(\tilde{x} + \int_0^T \varphi_s^*(u)\,dS(u))]\hspace{1cm}\Box$$
\normalfont
%%%%%%%%%%%%%%%%%%%%%%%%%%%%%%%%%%%%%%%%%%%%%%%%%%%%%%%%%%%%%%%%%%%%%%%%%%%%
\subsection{Final result on utility maximisation problem}
%%%%%%%%%%%%%%%%%%%%%%%%%%%%%%%%%%%%%%%%%%%%%%%%%%%%%%%%%%%%%%%%%%%%%%%%%%%%
We combine the results of Proposition \ref{t1} and Proposition \ref{t2}  to get the following final result on utility maximisation.

%%%%%%%%%%%%%%%%%%%%%%%%%%%%%%%%%%%%%%%%%%%%%%%%%%%%%%%%%%%%%%%%%%%%%%%%%%%%%%%
\thm \label{t3}
%%%%%%%%%%%%%%%%%%%%%%%%%%%%%%%%%%%%%%%%%%%%%%%%%%%%%%%%%%%%%%%%%%%%%%%%%%%%%%
We suppose that The Assumptions \ref{a1}, \ref{a11}, \ref{a2}, \ref{a3} hold. Then,
the maximal expected utility verify:
\begin{equation}
V(x,g)=\int_{\Xi}E_{P^u}\left[U\left(I\left(\lambda_g(u)\tilde{Z}^{*}_T(u)\right)\right)\right]d\alpha(u),
\label{64}
\end{equation}
 \begin{center}{and}\end{center}
\begin{equation}
V(x,0)=\int_{\Xi}E_{P^u}\left[U\left(I\left(\lambda_0(u)\tilde{Z}^{*}_T(u)\right)\right)\right]d\alpha(u),
\label{65}
\end{equation}
\normalfont
where $\lambda_g(u)$ is a solution of \eqref{62} and  $\lambda_0$ is a solution of \eqref{62} with replacing $g(u)$ by $0$.

%%%%%%%%%%%%%%%%%%%%%%%%%%%%%%%%%%%%%%%%%%%%%%%%%%%%%%%%%%%%%%%%%%%%%%%%%%%%%%

\end{section}
%%%%%%%%%%%%%%%%%%%%%%%%%%%%%%%%%%%%%%%%%%%%%%%%%%%%%%%%%%%%%%%%%%%%%%%%%%%%%%%
\begin{section}{Utility maximisation for HARA utilities} \label{s3a} %%%%%%%%%%%%%%%%%%%%%%%%%%
%%%%%%%%%%%%%%%%%%%%%%%%%%%%%%%%%%%%%%%%%%%%%%%%%%%%%%%%%%%%%%%%%%%%%%%%%%%%%%%

\par In the overwhelming part of the literature, the utility maximisation analysis is carried out under the hyperbolic absolute risk utilities (HARA), which are  logarithmic, power and exponential utilities represented below:
\begin{eqnarray*}
U(x) & =&\ln x, \text{ then} \,f(x)=-\ln{x}-1,\\
U(x) & =& \frac{x^p}{p},\ p\in(-\infty,0)\cup(0,1), \text{ then } f(x)= -\frac{p-1}{p}x^{\frac{p}{p-1}},\\
U(x) & =& 1-e^{-\gamma x},\ \gamma>0, \text{ then } f(x)=1-\frac{x}{\gamma}+\frac{1}{\gamma}x\ln{x}-\frac{1}{\gamma}x\ln{\gamma}
\end{eqnarray*}
where $x>0$. This choice can be explained by the good properties of these functions such as scaling property, time horizon invariance property, preservation of Levy property and so on (see for instance \cite{CV2}).
\par We introduce the information quantities related with HARA utilities. The corresponding maximal utilities are given in Theorem \ref{t4}. Then, we express these information quantities via information processes (cf. Propositions \ref{p1},
\ref{p2}, \ref{p3}). The final result on utility maximisation is given in Theorem \ref{t5}.
\subsection{Maximal utilities and information quantities}

\par As before, we assume the existence of $f$-divergence minimal martingale measure $Q^{u, *}\in\mathcal{K}^u.$
We introduce three important quantities related with $P^u_T$ and $Q^{u, *}_T$ namely the entropy of $P^{u}$ with respect to $Q^{u, *}_T$,
\begin{equation*}
{\bf{I}}(P^{u}_T|Q^{u,*}_T)=-E_{P^u}\left[\ln{\tilde{Z}_T^{*}(u)}\right],
\end{equation*}
the entropy of $Q^{u, *}_T$ with respect to $P^{u}_T$,
\begin{equation*}
{\bf{I}}(Q^{u, *}_T|P^{u}_T)=E_{P^u}\left[\tilde{Z}_T^{*}(u)\ln{\tilde{Z}_T^{*}(u)}\right],
\end{equation*}
and Hellinger type integrals 
\begin{equation*}
{\bf H}^{(q), *}_T(u)=E_{P^u}\left[(\tilde{Z}_T^{*}(u))^q\right], 
\end{equation*}
where $q=\frac{p}{p-1}$ and $ p<1.$
\par Now we give the expressions of the value function $V(x,g)$ involving relative entropies and Hellinger type integrals.
%%%%%%%%%%%%%%%%%%%%%%%%%%%%%%%%%%%%%%%%%%%%%%%%%%%%%%%%%%%%%%%%%%%%%%%%%%%%%
\thm \label{t4}
%%%%%%%%%%%%%%%%%%%%%%%%%%%%%%%%%%%%%%%%%%%%%%%%%%%%%%%%%%%%%%%%%%%%%%%%%%%%%%
Under The Assumptions \ref{a1}, \ref{a11}, \ref{a2}, \ref{a3} 
the information quantities are $\mathcal H$-measurable and  we have the following expressions for $V_T(x,0):$

$(i)$ If $U(x)=\ln{x}$  then 
\begin{equation}
V_T(x,0)=\int_{\Xi}[\,\ln{x}+{\bf{I}}(P^{u}_T|Q^{u, *}_T)\,]d\alpha(u)
\label{72a}
\end{equation}

$(ii)$ If $U(x)=\frac{x^p}{p}$ with $p<1, p\neq 0$ then
\begin{equation}
V_T(x,0)=\frac{1}{p}\int_{\Xi}x^p\left({\bf H}^{(q), *}_T(u)\right)^{1-p}d\alpha(u)
\label{73a}
\end{equation}

$(iii)$ If $U(x)=1-e^{-\gamma x}$ with $\gamma>0$  then
\begin{equation}
V_T(x,0)=1-\int_{\Xi}\,\exp \{-[\,\gamma x+{\bf I}(Q^{u, *}_T|P^{u}_T)\,]\}\,d\alpha(u)
\label{74a}
\end{equation}
The expressions for $V_T^u(x,g)$ can be obtained from previous expressions replacing in right-hand side $x$ by
$x+g(u)$.

\pf\normalfont
First of all we remark that  $\tilde{Z}_T^{*}$ is $\mathcal F_T\otimes \mathcal H$ measurable and  $\frac{dP^u_T}{dP_T}=p_T$ is also $\mathcal F_T\otimes \mathcal H$ measurable. Hence, the information quantities are  $\mathcal H$-measurable.
\par $(i)$ The Theorem \ref{t3} states that
\begin{equation}
V_T(x,0)=\int_{\Xi}E_{P^u}\left[U\left(I\left(\lambda_0(u)\,\tilde{Z}^{*}_T(u)\right)\right)\right]d\alpha(u), 
\label{21}
\end{equation}
where $\lambda_0(u)$ is defined from the equation
\begin{eqnarray}
E_{P^u}\left[\tilde{Z}_T^{*}(u)I\left(\lambda_0(u)\tilde{Z}_T^{*}(u)\right)\right] = x.
\label{24}
\end{eqnarray}
The corresponding inverse function of the derivative of the logarithmic utility is  $I(y) =\frac{1}{y}$, then $\lambda_0(u)=\frac{1}{x}.$
Putting this result into \eqref{21} we have that
\begin{eqnarray}
V_T(x,0)&=&\int_{\Xi}E_{P^u}\left[\ln\left[\frac{x}{\tilde{Z}^{*}_T(u)}\right]\right]d\alpha(u)\nonumber\\
&=&\int_{\Xi}E_{P^u}\left[\ln{x}-\ln{\tilde{Z}^{*}_T(u)}\right]d\alpha(u)\nonumber\\
&=&\int_{\Xi}[\,\ln{x}+{\bf{I}}(P^{u}_T|Q^{u, *}_T)\,]d\alpha(u).
\label{28}
\end{eqnarray}
The formula \eqref{72} has been proved.

$(ii)$ 
The corresponding inverse function of the derivative of the power utility is
$I(y) = y^{\frac{1}{p-1}}$. Then, using  \eqref{24} we deduce that
\begin{eqnarray*}
\lambda_0(u)=\frac{x^{p-1}}{\left(E_{P^u}\left[\left(\tilde{Z}_T^{*}(u)\right)^{q}\right]\right)^{p-1}}.
\end{eqnarray*}
with $q=\frac{p}{p-1}$, and  we get finally that
\begin{eqnarray}
E_{P^u}\left[U\left(I\left(\lambda_0(u)\,\tilde{Z}^{*}_T(u)\right)\right)\right]&=&E_{P^u}\left[\frac{1}{p}\left(\lambda_0(u)\,\tilde{Z}^{*}_T(u)\right)^{q}\right]\nonumber\\
&=&\frac{x^p}{p}\left(E_{P^u}\left[\left(\tilde{Z}_T^{*}(u)\right)^{q}\right]\right)^{1-p}.
\label{32}
\end{eqnarray}
Then, we integrate over $\Xi$ with respect to $\alpha$ and we obtain \eqref{73a}.

$(iii)$ The corresponding inverse function of the derivative of the exponential utility is $I(y)= -\frac{1}{\gamma}\left(\ln{y}-\ln{\gamma}\right)$.
Then, value function in the case of the exponential utility can be simplified to the form
\begin{eqnarray}
V_T(x,0)&=&\int_{\Xi}E_{P^u}\left[U\left(I\left(\lambda_{0}(u)\tilde{Z}_T^{*}(u)\right)\right)\right]d\alpha(u)\nonumber\\
&=&1-\frac{1}{\gamma} \int_{\Xi}\lambda_0(u)\,d\alpha(u).
\end{eqnarray}
The corresponding  $\lambda_0$ is given by
\begin{eqnarray*}
\lambda_0(u)=\gamma\exp\bigg\{-\gamma x-E_{P^u}\left[\tilde{Z}_T^{*}(u)\ln{\tilde{Z}_T^{*}(u)}\right]\bigg\}.
\end{eqnarray*}
Taking into account that $$E_{P^u}\left[\tilde{Z}_T^{*}(u)\ln{\tilde{Z}_T^{*}(u)}\right]={\bf{I}}(Q^{u, *}_T|P^{u}_T)$$ we get  that
\begin{eqnarray*}
\lambda_0(u)=\gamma\exp\bigg\{-\gamma x - {\bf{I}}(Q^{u, *}_T|P^{u}_T)\bigg\}.
\end{eqnarray*}
and it gives us \eqref{74a}. 
$\Box$

\subsection{Information quantities and information processes}

\par In this subsection we express the information quantities via corresponding information processes. As previously, we assume the existence of an equivalent $f$-divergence minimal martingale measure $Q^{u, *}$. We recall that a semi-martingale $X(\xi)$ under $P^u$ is also a semi-martingale with the triplet $T^{{\bf F}}(u)=(B^u,C,\nu^u)$ defined by \eqref{5}. To avoid non-necessary complications we introduce the following additional assumption.
\asm \label{a4} For each $u\in \Xi$, the $(P^u,{\bf F})$-semi-martingale $X$ is a quasi-left continuous, i.e. for any predictable stopping time $\tau$, the jump $\Delta X_{\tau}=0$ on the set  $\{\tau <\infty \}$. 
\normalfont
\par Let us denote by $\beta^{u, *}$ and $Y^{u, *}(x)$ two $(P^u,{\bf F})$-predictable processes known as Girsanov parameters for the changing of measure from $P^u$ into $Q^{u, *}$ such that: $\forall t\geq 0$ and $P^u$-a.s.
\begin{equation*}
\int_0^t\int_{\mathbb{R}}|\,l(x)\,(Y_s^{u,*}(x)-1)|\nu^u(ds,dx)<\infty,\ \ \int_0^t(\beta_s^{u,*})^2dC_s<\infty.
\end{equation*}
In the case of logarithmic utility we consider the entropy ${\bf{I}}(P^{u}_T\,|\,Q^{u, *}_t)$ and we introduce the corresponding predictable process $\mathcal{I}^{*}(u)=(\mathcal{I}^{*}_t(u))_{t\in\left[0,T\right]}$
\begin{equation}
\mathcal{I}^{*}_t(u)=\frac{1}{2}\int_0^t(\beta_s^{u, *})^2dC_s-\int_0^t\int_{\mathbb{R}}\left(\ln(Y_s^{u, *}(x))-Y_s^{u, *}(x)+1\right)\nu^u(ds,dx).
\label{71}
\end{equation}
%%%%%%%%%%%%%%%%%%%%%%%%%%%%%%%%%%%%%%%%%%%%%%%%%%%%%%%%%%%%%%%%%%%%%%%%%%
\prop \label{p1}
%%%%%%%%%%%%%%%%%%%%%%%%%%%%%%%%%%%%%%%%%%%%%%%%%%%%%%%%%%%%%%%%%%%%%%%%%%
We suppose  that $E_{P^u}|\ln{\tilde{Z}_T^{*}(u)}|<\infty$ and Assumption \ref{a4}
holds. Then 
\begin{equation}
{\bf{I}}(P^{u}_T\,|\,Q^{u, *}_T)=E_{P^u}\mathcal{I}^{*}_T(u).
 \label{76}
 \end{equation}

\pf\normalfont
To avoid the complicated notations we omit for the proof of this Proposition the indexes $u$, ${*}$, and replace the notation of $\tilde{Z}$ by $Z$. Let $Q$ and $P$ be two equivalent probability  measures on canonical space and let $(Z_t)_{t\in\left[0,T\right]}$ be the Radon-Nikodym density, $Z_t=\frac{dQ_t}{dP_t}$, where $Q_t$ and $P_t$ are the restrictions of $Q$ and $P$ on $\sigma$-algebra $\mathcal{F}_t$. Let $X$ be $P$-semi-martingale with the characteristics $(B,C,\nu).$ 
\par For $\epsilon>0$ we put 
\begin{equation}
\label{tau}
\tau_\epsilon=\inf\{0\leq  t\leq T| Z_t\leq \epsilon\} 
\end{equation}
with $\inf\{\emptyset\}=+\infty$. We remark that  $\tau_\epsilon$ is a stopping time and that
the sequence of stopping times  $(\tau_\epsilon)$  is increasing to infinity as $\epsilon\rightarrow 0$, and hence, it is localising sequence. Then, by Ito formula we have:
\begin{multline}
\ln{Z}_{T\wedge\tau_{\epsilon}}=\ln{Z}_0+\int_0^{T\wedge\tau_{\epsilon}}\frac{1}{Z_{s-}}dZ_s-\frac{1}{2}\int_0^{T\wedge\tau_{\epsilon}}\frac{1}{(Z_{s-})^2}d<Z^c>_s\\
+\underset{0<s\leq T\wedge \tau_{\epsilon}}{\sum}\left(\ln{Z_s}-\ln{Z_{s-}}-\frac{1}{Z_{s-}}\Delta Z_s\right),
\label{77}
\end{multline}
where $\Delta Z_s=Z_s-Z_{s-}$ and $<Z^c>_s$ is a predictable variation of the continuous martingale part of $Z$. We remark that $\left(\int_0^{t\wedge\tau_{\epsilon}}\frac{1}{Z_{s-}}dZ_s\right)_{t\in\left[0,T\right]}$ is a $(P,{\bf F})$-martingale started from zero, since it is stochastic integral with respect to $(P,{\bf F})$-martingale $Z$  and since $Z_{s-}\geq\epsilon>0$ on the stochastic interval $\left[0,T\wedge\tau_{\epsilon}\right].$ By Theorem 1.8, p.66 in \cite{JSh}, we get 
\begin{multline*}
E_P\int_0^{t\wedge{\tau_{\epsilon}}}\int_{\mathbb{R}}\left(\ln{\left(1+\frac{x}{Z_{s-}}\right)-\frac{x}{Z_{s-}}}\right)\mu_Z(ds,dx)=\\
E_P\int_0^{t\wedge{\tau_{\epsilon}}}\int_{\mathbb{R}}\left(\ln{\left(1+\frac{x}{Z_{s-}}\right)-\frac{x}{Z_{s-}}}\right)\nu_Z(ds,dx),
\end{multline*}
where $\mu_Z$ and $\nu_Z$ are measure of jumps  of $Z$ and its compensator.
Finally, from \eqref{77} and the fact that $Z_0=1$, we have:
\begin{multline}
E_P\ln{Z}_{T\wedge\tau_{\epsilon}}=E_P\left[-\frac{1}{2}\int_0^{T\wedge\tau_{\epsilon}}\frac{1}{(Z_{s-})^2}d<Z^c>_s\right.\\
\left.+\int_0^{T\wedge{\tau_{\epsilon}}}\int_{\mathbb{R}}\left(\ln{\left(1+\frac{x}{Z_{s-}}\right)-\frac{x}{Z_{s-}}}\right)\nu_Z(ds,dx)\right].
\label{79}
\end{multline} 
\par Now, since $Z=\mathcal{E}(M)$ where $\mathcal{E}(\cdot)$ is a Dolean-Dade exponential, for all $t\in\left[0,T\right]$ we get:
\begin{equation}
M_t=\int_0^t\beta_sdX_s^c+\int_0^t\int_{\mathbb{R}}(Y_s-1)(\mu-\nu)(ds,dx)
\label{78}
\end{equation}
Then, $dZ_t=Z_{t-}dM_t,$ and in particular, $dZ_t^c=Z_tdM_t^c$ and $\Delta Z_t=Z_{t-}\Delta M_t.$ In addition, \eqref{78} implies that for $t\in\left[0,T\right]$
\begin{equation}
M_t^c=\int_0^t\beta_sdX_s^c\text{ and } \Delta M_t=\left(Y_t(\Delta X_t)-1\right),
\label{86}
\end{equation}
and, hence, 
$$d<Z^c>_t=(Z_{t-})^2\beta_t^2d<X^c>_t$$
 and
$$\Delta Z_t=Z_{t-}\left(Y_t(\Delta X_t)-1\right).$$
Using mentioned above relations we obtain:
\begin{equation}
E_P\ln{Z}_{T\wedge\tau_{\epsilon}}=E_P\left[-\frac{1}{2}\int_0^{T\wedge\tau_{\epsilon}}\beta_s^2dC_s+\int_0^{T\wedge{\tau_{\epsilon}}}\int_{\mathbb{R}}\left(\ln{Y_s(x)}-Y_s(x)+1\right)\nu(ds,dx)\right].
\label{79a}
\end{equation} 
Since $\ln(1+x)\leq x$, both integrands in the right hand side are negatives. So, using the Lebesgue monotone convergence theorem, we can pass to the limit on the right-hand side.
\par It remains to pass to the limit on the left hand side in \eqref{79a}, i.e. to prove 
\begin{equation}\label{lim}
\lim_{\epsilon\rightarrow 0}\,E_P\ln{Z}_{T\wedge\tau_{\epsilon}}=E_P \ln Z_T.
\end{equation}
We can write 
\begin{equation}
E_P\ln{Z}_{T\wedge\tau_{\epsilon}} - E_P\ln Z_T= E_P\left[\ln Z_{\tau_{\epsilon}} 1_{\{{\tau_{\epsilon}}<T\}}\right]- E_P\left[\ln Z_T 1_{\{{\tau_{\epsilon}}< T\}}\right].
\label{203a}
\end{equation}
We show that two last terms in \eqref{203a} tend to zero as $\epsilon\rightarrow 0$. 
\par Let $\hat{Z}_t = \frac{1}{Z_t}$ for $t\in [0,T]$. We remark that $(\hat{Z}_t)_{[0,T]}$ is a $Q$-martingale.  Then, by maximal inequality for positive martingales
\begin{equation}
Q\left(\tau_{\epsilon}< T\right)\leq Q\left(\underset{0\leq t\leq T}{\sup}\hat{Z}_t\geq\frac{1}{\epsilon}\right)\leq E_Q \hat{Z}_T \cdot\epsilon=\epsilon
\label{204}
\end{equation}
Finally,
$$P\left(\tau_{\epsilon}< T\right)\leq E_Q(\hat{Z}_T 1_{\{ \underset{0\leq t\leq T}{\sup}\hat{Z}_t\geq\frac{1}{\epsilon}\}})\rightarrow 0$$
as $\epsilon \rightarrow 0$ since $E_Q \hat{Z}_T = 1$.
Since $\ln Z_T$ is $P$-integrable, the relation \eqref{204} implies that
\begin{equation}\label{204a}
\lim_{\epsilon\rightarrow 0}E_P\left(\ln{Z}_{T}1_{\{\tau_{\epsilon}< T\}}\right)=0.
\end{equation} 
\par Since $Z_{\tau_{\epsilon}}\leq \epsilon$, for  
$\epsilon <1$ we get $E_P(\ln Z_{\tau_{\epsilon}}1_{\{\tau_{\epsilon}< T\}}) \leq 0$.  From concavity of $\ln x, x>0$
\begin{equation*}
E_P(\ln Z_{\tau_{\epsilon}}1_{\{\tau_{\epsilon}< T\}})\geq E_P\left(\ln{Z}_{T}1_{\{\tau_{\epsilon}< T\}}\right)
\end{equation*}
The relation \eqref{204a} implies that $E_P\left(\ln{Z}_{\tau_{\epsilon}}1_{\{\tau_{\epsilon}< T\}}\right)\rightarrow 0$ when $\epsilon\rightarrow 0$.
 Finally, we proved (\ref{lim}).
$\Box$

\par In the case of exponential utility we consider the entropy ${\bf{I}}(\,Q^{u,*}_T\,|\,P^{u}_T\,)$, which is known also as Kullback-Leiber information. We introduce the corresponding Kullback-Leiber process ${I}^{*}(u)=({I}^{*}_t(u))_{t\in\left[0,T\right]}$ with
\begin{equation}{I}^{*}_t(u)=\frac{1}{2}\int_0^t(\beta_s^{u, *})^2dC_s+\int_0^t\int_{\mathbb{R}}\left[Y_s^{u, *}(x)\ln(Y_s^{u, *}(x))-Y_s^{u, *}(x)+1\right]\nu^u(ds,dx).
\label{83}
\end{equation}
We remark that, Kullback-Leibler process was first introduced  in \cite{Kol} and it was studied in \cite{M}, \cite{ES}, \cite{HS}. We give here some properties of this process needed for our final results.
%%%%%%%%%%%%%%%%%%%%%%%%%%%%%%%%%%%%%%%%%%%%%%%%%%%%%%%%%%%%%%%%%%%%%%%%%%%%%%%
\prop \label{p2} 
%%%%%%%%%%%%%%%%%%%%%%%%%%%%%%%%%%%%%%%%%%%%%%%%%%%%%%%%%%%%%%%%%%%%%%%%%%%%%%

We suppose that $E_{P^u}|{\tilde{Z}_T^{*}(u)}\ln{\tilde{Z}_T^{*}(u)}|<\infty$ and that the Assumption \ref{a4} holds. Then, 
\begin{equation}
{\bf{I}}(\,Q^{u, *}_T\,|\,P^{u}_T\,)=E_{P^u}\left[\int_0^T\tilde{Z}_{s-}^{*}(u)\,d{I}^{*}_s(u)\right] = E_{Q^{u,*}}\left(\,{I}^{*}_T(u)\,\right) 
 \label{83a}
 \end{equation}
 
\pf\normalfont We continue in the framework of Proposition \ref{p1} to prove the first equality. The second one is a consequence of integration by part formula.
Let $\epsilon>0$ and the localising sequence $(\tau_{\epsilon})$ : for $\epsilon >0$
$$\tau_\epsilon=\inf\{0\leq  t\leq T| Z_t\leq \epsilon\,\mbox{or}\,Z_t\geq \frac{1}{\epsilon} \}$$
with $\inf \{\emptyset \}= + \infty$.
Then, by Ito formula we have
\begin{multline}
{Z}_{T\wedge\tau_{\epsilon}}\ln Z_{T\wedge\tau_{\epsilon}}={Z}_0\ln{Z}_0+\int_0^{T\wedge\tau_{\epsilon}}(\ln Z_{s-}+1)dZ_s+\frac{1}{2}\int_0^{T\wedge\tau_{\epsilon}}\frac{1}{Z_{s-}}d<Z^c>_s\\
+\int_0^{T\wedge\tau_{\epsilon}}\int_{\mathbb{R}}\left[(Z_{s-}+x)\ln (Z_{s-}+x)-Z_{s-}\ln Z_{s-}-(\ln Z_{s-}+1)x\right]\mu_Z(ds,dx).
\label{84}
\end{multline}
We remark that $\left(\int_0^{T\wedge\tau_{\epsilon}}(\ln Z_{s-}+1)dZ_s\right)_{t\in\left[0,T\right]}$ is a $(P,{\bf F})$-martingale started from zero, since it is stochastic integral with respect to $(P,{\bf F})$-martingale $Z$  such that  $Z_{s-}\geq\epsilon $  and $Z_{s-}\leq \frac{1}{\epsilon} $ on the stochastic interval $\left[0,T\wedge\tau_{\epsilon}\right]$. Using  Theorem 1.8, p.66, in \cite{JSh}, we get 
\begin{multline*}
E_P\int_0^{t\wedge{\tau_{\epsilon}}}\int_{\mathbb{R}}\left[(Z_{s-}+x)\ln (Z_{s-}+x)-Z_{s-}\ln Z_{s-}-(\ln Z_{s-}+1)x\right]\mu_Z(ds,dx)=\\
E_P\int_0^{t\wedge{\tau_{\epsilon}}}\int_{\mathbb{R}}\left[(Z_{s-}+x)\ln (Z_{s-}+x)-Z_{s-}\ln Z_{s-}-(\ln Z_{s-}+1)x\right]\nu_Z(ds,dx),
\end{multline*}
where $\mu_Z$ and $\nu_Z$ are the measure of jumps of $Z$ and its compensator.
\par Finally, from \eqref{84} and  the fact that $Z_0=1$, we have:
\begin{multline}
E_P\left[Z_{T\wedge\tau_{\epsilon}}\ln Z_{T\wedge\tau_{\epsilon}}\right]=E_P\left[\frac{1}{2}\int_0^{T\wedge\tau_{\epsilon}}\frac{1}{Z_{s-}}d<Z^c>_s\right.\\
\left.+\int_0^{T\wedge{\tau_{\epsilon}}}\int_{\mathbb{R}}\left[(Z_{s-}+x)\ln (Z_{s-}+x)-Z_{s-}\ln Z_{s-}-(\ln Z_{s-}+1)x\right]\nu_Z(ds,dx)\right].
\label{85}
\end{multline} 
Using the relations between  $Z=\mathcal{E}(M)$, $M$ and the process $X$ given by \eqref{78} and \eqref{86}, we get
\begin{multline}
E_P\left[Z_{T\wedge\tau_{\epsilon}}\ln Z_{T\wedge\tau_{\epsilon}}\right]=E_P\left[\frac{1}{2}\int_0^{T\wedge\tau_{\epsilon}}Z_{s-}\beta^2_sdC_s\right.\\
\left.+\int_0^{T\wedge{\tau_{\epsilon}}}\int_{\mathbb{R}}Z_{s-}\left(Y_s(x)\ln Y_s(x)-Y_s(x)+1\right)\nu(ds,dx)\right].
\label{87}
\end{multline} 
\par Since $\tau_{\epsilon}\rightarrow +\infty$ as $\epsilon\rightarrow 0$  and $x\ln x-x+1\geq 0 $ for all $x>0$, by Lebesgue monotone convergence theorem  we can pass to the limit in the right-hand side of \eqref{87}. It remains to show that the left-hand side of \eqref{87} converges to $E_P\left[Z_T\ln Z_T\right].$ 
We can write 
\begin{equation}\label{80}
E_P\left[{Z}_{T\wedge\tau_{\epsilon}}\ln{Z}_{T\wedge\tau_{\epsilon}}\right]-E_P\left[{Z}_{T}\ln{Z}_{T}\right]=
\end{equation}
$$E_P\left[{Z}_{\tau_{\epsilon}}\ln{Z}_{\tau_{\epsilon}}1_{\{\tau_{\epsilon}< T\}}\right]-E_P\left[{Z}_{T}\ln{Z}_{T}1_{\{\tau_{\epsilon}< T\}}\right].
$$
\par We show that the last two terms in \eqref{80} tends to zero as $\epsilon\rightarrow 0$. Since ${Z}_{T}\ln{Z}_{T}$ is $P$-integrable and  $P\left(\tau_{\epsilon}<T\right)\rightarrow 0$ as $\epsilon\rightarrow 0$, we get that 
\begin{equation*}
\lim_{\epsilon\rightarrow 0}E_P\left[{Z}_{T}\ln{Z}_{T}1_{\{\tau_{\epsilon}< T\}}\right]=0.
\end{equation*} 
\par Using the inequality $x\ln{x}\geq\frac{1}{e}$ for all $x\in\mathbb{R}_{+},$ we have for $0\leq\epsilon\leq\frac{1}{e}$ 
\begin{equation*}
-\frac{1}{e}\cdot P\left(\tau_{\epsilon}<T\right)\leq E_P\left[{Z}_{\tau_{\epsilon}}\ln{Z}_{\tau_{\epsilon}}1_{\{\tau_{\epsilon}< T\}}\right]\leq 0,
\end{equation*}
and $E_P\left[{Z}_{\tau_{\epsilon}}\ln{Z}_{\tau_{\epsilon}}1_{\{\tau_{\epsilon}< T\}}\right]\rightarrow 0$ when $\epsilon\rightarrow 0$.
Finally, 
$$\underset{\epsilon\rightarrow 0}{\lim}E_P\left[{Z}_{T\wedge\tau_{\epsilon}}\ln{Z}_{T\wedge\tau_{\epsilon}}\right]=E_P\left[{Z}_{T}\ln{Z}_{T}\right].$$
and the proposition is proved.
$\Box$
\par For the case of power utility we consider Hellinger types integrals 
$${\bf H}^{(q), *}_T(u)=E_{P^u}\left[(\tilde{Z}_T^{*}(u))^q\right],$$
where $ q=\frac{p}{p-1},\ \ p<1$. We notice that if $p<0$ then $0<q<1$ and if $0<p<1$ then $q<0$, so, in any cases  $q<1$.
 \par We introduce the corresponding predictable process called Hellinger type process $h^{(q), *}(u)=(h^{(q), *}_t(u))_{t\in\left[0,T\right]}$
\begin{equation}\label{88}
h^{(q), *}_t(u)=\frac{1}{2}q(q-1)\int_0^t(\beta_s^{u, *})^2dC_s+
\end{equation}
$$\int_0^t\int_{\mathbb{R}}\left[\left(Y_s^{u, *}(x)\right)^{q}-q(Y_s^{u, *}(x)-1) -1\right]\nu(ds,dx),$$
In the case when $0<q<1$ the Hellinger processes was studied in \cite {VV}, \cite{JSh}, \cite{CS},\cite{CSL}. We show that the result can be extended for $q<1$.
%%%%%%%%%%%%%%%%%%%%%%%%%%%%%%%%%%%%%%%%%%%%%%%%%%%%%%%%%%%%%%%%%%%%%%%%%%%%%%%%%%%%%%%%%
\prop \label{p3} Suppose that ${\bf H}^{(q), *}_T(u)< \infty$ and that the Assumption \ref{a4} holds. Then
\begin{equation*}
{\bf H}^{(q), *}_T(u)=1+E_{P^u}\left[\int_0^T(\tilde{Z}_{s-}^{*})^q\,dh_s^{(q),*}(u)\right]
\end{equation*}
or, in the terms of the stochastic exponential:
\begin{equation}
{\bf H}^{(q), *}_T(u)=E_{P^u}\left[\mathcal{E}\left(h^{(q), *}\right)_T\right].
\label{hel}
\end{equation}
%%%%%%%%%%%%%%%%%%%%%%%%%%%%%%%%%%%%%%%%%%%%%%%%%%%%%%%%%%%%%%%%%%%%%%%%%%%%%%%%%%%%%%%%%

\pf\normalfont We continue in the framework of Proposition \ref{p1}.
Let $\epsilon>0$ and the localising sequence $(\tau_{\epsilon})$ defined by (\ref{tau}). Applying Ito's formula we have:
\begin{eqnarray*}
Z^{q}_{T\wedge \tau_{\epsilon}}&=&1+q\int_0^{T\wedge \tau_{\epsilon}}Z_{s-}^{q-1}dZ_s+\frac{1}{2}q(q-1)\int_0^{T\wedge \tau_{\epsilon}}Z_{s-}^{q-2}d<Z^c>_{s}\\
&+&\int_0^{T\wedge \tau_{\epsilon}}\int_{\mathbb{R}}\left[Z_{s-}^{q}\left(\left(1+\frac{x}{Z_{s-}}\right)^q -1\right)-qZ_{s-}^{q-1}x\right]\mu_Z(ds,dx).
\end{eqnarray*}
Since $\left(\int_0^{t\wedge \tau_{\epsilon}}Z_{s-}^{q-1}dZ_s\right)_{t\in\left[0,T\right]}$  is a $(P,{\bf F})$-martingale starting from $0$ and due to the projection theorem we get:
\begin{eqnarray*}
E_{P}Z^{q}_{T\wedge \tau_{\epsilon}}&=&1+E_P\left[\frac{1}{2}q(q-1)\int_0^{T\wedge \tau_{\epsilon}}Z_{s-}^{q-2}d<Z^c>_{s}\right.\\
&+&\left.\int_0^{T\wedge \tau_{\epsilon}}\int_{\mathbb{R}}\left[Z_{s-}^{q}\left(\left(1+ \frac{x}{Z_{s-}}\right)^q-1\right)-qZ_{s-}^{q-1}x\right]\nu_Z(ds,dx)\right].
\end{eqnarray*}
\par Using the relation between $Z=\mathcal{E}(M)$, $M$ and the initial process $X$, we get
\begin{eqnarray*}
E_{P}
Z^{q}_{T\wedge \tau_{\epsilon}}&=&1+E_P\left\{\frac{1}{2}q(q-1)\int_0^{T\wedge \tau_{\epsilon}}Z_{s-}^{q}\beta_s^2dC_s\right.\\
&+&\left.\int_0^{T\wedge \tau_{\epsilon}}\int_{\mathbb{R}}Z_{s-}^{q}\left[Y^q_s(x)-q(Y_s(x)-1)-1\right]\nu (ds,dx)\right\}
\end{eqnarray*}
We remark that $\underset{\epsilon\rightarrow 0}{\lim}\tau_{\epsilon}= +\infty$. Since for $0<q<1$, $q(q-1)<0$ and $x^q-qx-1\leq 0$ and for $q<0$, $q(q-1)>0$ and $x^q-qx-1\geq 0$, the right hand side of above expression contains the integral of some negative function. Then, by Lebesgue monotone convergence theorem we can pass to the limit on the right hand side.
\par It remains to show that
\begin{equation*}
\underset{\epsilon\rightarrow 0}{\lim} E_{P}
Z^{q}_{T\wedge \tau_{\epsilon}}=E_P Z_T^q.
\end{equation*}
We have:
\begin{equation}
E_{P}Z^{q}_{T\wedge \tau_{\epsilon}}-E_P Z_T^q=E_P\left(Z_{\tau_{\epsilon}}^q 1_{\{\tau_{\epsilon}<T\}}\right)-E_P\left(Z_T^q 1_{\{\tau_{\epsilon}<T\}}\right).
\label{90}
\end{equation}
\par Since  $P\left(\tau_{\epsilon}<T\right)\rightarrow 0$ as $\epsilon\rightarrow 0$ and $Z_T^q$ is $P$-integrable, the first term in the right-hand side of \eqref{90} tends to zero. For the second term we distinguish two cases: $0<q<1$ and
$q<0$. In the first case, 
$$E_P\left(Z_{\tau_{\epsilon}}^q 1_{\{\tau_{\epsilon}<T\}}\right) \leq \epsilon ^q 
P\left(\tau_{\epsilon}<T\right)\rightarrow 0$$
as $\epsilon\rightarrow 0$.
In the second case we have:
\begin{equation*}
E_P\left(Z_{\tau_{\epsilon}}^q 1_{\{\tau_{\epsilon}<T\}}\right)=E_Q\left(\hat{Z}_{\tau_{\epsilon}}^{1-q} 1_{\{\tau_{\epsilon}<T\}}\right),
\end{equation*} 
where $\hat{Z}_{\tau_{\epsilon}}=\displaystyle\frac{1}{Z_{\tau_{\epsilon}}}.$ 
From maximal inequalities for the martingales we have:
 $$E_Q\left[\underset{0\leq t\leq T}{\sup}\hat{Z}_t\right]^{1-q} \leq c(q) E_Q (\hat{Z}_T^{1-q}) = c(q) E_P(Z_T^q )<\infty$$
 where $c(q)$ is a constant.
In addition, $Q\left(\tau_{\epsilon}<T\right)\rightarrow 0$ as $\epsilon\rightarrow 0$, and, then,
\begin{equation*}
\underset{\epsilon\rightarrow 0}{\lim}E_Q\left(\hat{Z}_{\tau_{\epsilon}}^{1-q} 1_{\{\tau_{\epsilon}<T\}}\right)=0.
\end{equation*}
We prove now \eqref{hel}. From Ito formula we also get that
$$Z^{q}_{T\wedge \tau_{\epsilon}} = 1+ \int_0^{T\wedge \tau_{\epsilon}} Z_s^q \,d K_s$$
where $K= N+h^{(q)}$ is a sum of a martingale $N$ and predictable process $h^{(q)}$
with
$$ N_t = q\int_0^t\beta_{s-}\,dX^c_s + \int_0^{t}\int_{\mathbb R}(Y^q_s(x)-1)(\mu -\nu)(ds, dx)$$
Then, we have:
$$Z^{q}_{T\wedge \tau_{\epsilon}}= \mathcal E(N+h^{(q)})_{T\wedge \tau_{\epsilon}}=
\mathcal E(N)_{T\wedge \tau_{\epsilon}}\,\mathcal E(h^{(q)})_{T\wedge \tau_{\epsilon}}$$
We take the expectation with respect to $P$ and we show that $(\mathcal E (N)_{t\wedge \tau_{\epsilon}})_{0\leq t\leq T}$ is uniformly integrable martingale
with expectation 1.
Since $h^{(q)}$ is monotone and continuous process, we can pass to the limit and it gives \eqref{hel}.
$\Box$

\subsection{Maximal utility and information processes}

The final result for maximal utility in terms of information processes  follows directly from Theorem \ref{t4} and Propositions \ref{p1}, \ref{p2} and \ref{p3}
and is given in the following Theorem \ref{t5}.
%%%%%%%%%%%%%%%%%%%%%%%%%%%%%%%%%%%%%%%%%%%%%%%%%%%%%%%%%%%%%%%%%%%%%%%%%%%%%
\thm \label{t5}
%%%%%%%%%%%%%%%%%%%%%%%%%%%%%%%%%%%%%%%%%%%%%%%%%%%%%%%%%%%%%%%%%%%%%%%%%%%%%
Under The Assumptions \ref{a1}, \ref{a11}, \ref{a2}, \ref{a3}, \ref{a4}  and for HARA utilities we have the following expressions for $V_T(x,0):$

$(i)$ If $U(x)=\ln{x}$, then 

\begin{equation}
V_T(x,0)=\int_{\Xi}E_{P^u}[\,\ln{x}+\mathcal{I}^{*}_T(u)]\,d\alpha(u)
\label{72}
\end{equation}

$(ii)$ If $U(x)=\frac{x^p}{p}$ with $p<1,p\neq 0$, then

\begin{equation}
V_T(x,0)=\frac{1}{p}\int_{\Xi}x^p\left(E_{P^u}\left[\mathcal{E}\left(h^{(q), *}(u)\right)_T\right]\right)^{1-p}d\alpha(u)
\label{73}
\end{equation}

$(iii)$ If $U(x)=1-e^{-\gamma x}$ with $\gamma>0$, then

\begin{equation}
V_T(x,0)=1-\int_{\Xi} \exp \{-(\gamma x+E_{Q^{u,*}}(\,{I}^{*}_T(u)\,)\}\,d\alpha(u)
\label{74}
\end{equation}
The expressions for $V_T(x,g)$ can be obtained from previous expressions replacing $x$ by
$x+g(u)$ in right-hand side.
\normalfont
\end{section}

%%%%%%%%%%%%%%%%%%%%%%%%%%%%%%%%%%%%%%%%%%%%%%%%%%%%%%%%%%%%%%%%%%%%%%%%%%%%%%
\begin{section}{Indifference pricing  on the initially enlarged filtration} \label{s4}
%%%%%%%%%%%%%%%%%%%%%%%%%%%%%%%%%%%%%%%%%%%%%%%%%%%%%%%%%%%%%%%%%%%%%%%%%%%%%%

We consider the situation when  the investor carries out the trading  of risky asset $S(\xi)$ on the finite time interval $\left[0,T\right]$ and has a European type option with the pay-off function $G_T=g(\xi)$,  $g$ is an $\mathcal{H}$-measurable real-valued function. Then, as it was already mentioned a buyer's indifference price $p^b_T$ is the solution to the equation
\begin{equation}
V_T(x,0)=V_T(x-p_T^b, g).
\label{102}
\end{equation}
and a seller's indifference price $p^s_T$ is defined from
\begin{equation}
V_T(x,0)=V_T(x+p_T^s, -g).
\label{102a}
\end{equation}
We notice that the indifference prices $p_T^b$ and $p^s_T$  are related, namely
\begin{equation}\label{pbps}
p_T^{b}(g)=-p_T^s(-g).
\end{equation}

\subsection{Indifference price formulas}

\par Now we apply the results of  Theorem \ref{t3} and Theorem \ref{t4}  to give the formulas for the indifference prices in the cases of the exponential, power and logarithmic utilities. 

%%%%%%%%%%%%%%%%%%%%%%%%%%%%%%%%%%%%%%%%%%%%%%%%%%%%%%%%%%%%%%%%%%%%%%%%%%%%%
\prop \label{p11}
%%%%%%%%%%%%%%%%%%%%%%%%%%%%%%%%%%%%%%%%%%%%%%%%%%%%%%%%%%%%%%%%%%%%%%%%%%%%%
In the case of logarithmic utility $U(x)=\ln x,\,x>0,$ and under  the Assumptions \ref{a1}, \ref{a11}, \ref{a2}, \ref{a3},  and  $g(\xi)\in ]0,x[$ ($\alpha$-a.s.),  the buyer's and seller's indifference price satisfy:
\begin{equation}
\int_{\Xi}\ln \left[1-\frac{p^{b}_T}{x}+\frac{g(u)}{x}\right]d\alpha(u)=0
\label{29}
\end{equation} 
and 
\begin{equation}
\int_{\Xi}\ln \left[1+\frac{p^{s}_T}{x}-\frac{g(u)}{x}\right]d\alpha(u)=0.
\label{290}
\end{equation} 
Moreover, if $\ln(g(\xi))$, $\ln (x-g(\xi))$ are integrable functions  then the solutions of the equations \eqref{29}  and  \eqref{290} exist, they are unique and  $p_T^b, p^s_T\in\left[0,x\right]$.

\rem It should be noticed that in logarithmic utility case, the formulas for indifference price do not reflect the dependence between $X(\xi)$ and $\xi$:
exactly the same equations will hold  when $X(\xi)$ and $\xi$ are independent.\rm

\pf\normalfont
From \eqref{72a} and \eqref{102}  we have
 \eqref{29}. Formula \eqref{290} is obtained from the relation \eqref{pbps}.
We see that 
\begin{equation*}
F(y)=\int_{\Xi}\ln \left[1-\frac{y}{x}+\frac{g(u)}{x}\right]d\alpha(u),\ \ y\in\left[0,x\right],
\end{equation*}
is well-defined strictly decreasing function and  $F(0)\geq 0.$ If $\ln g(\xi)$ is integrable with respect to $\alpha$, then $F$ is a continuous by Lebesgue  dominated theorem. Under the condition that   $g(\xi)\in ]0,x[$ ($\alpha$-a.s.), $F(x)\leq 0$. Then, solution exists by the mean-value theorem and it is unique. 
\par In the case of the seller's indifference price, the function 
\begin{equation*}
F(y)=\int_{\Xi}\ln \left[1+\frac{y}{x}-\frac{g(u)}{x}\right]d\alpha(u),\ \ y\in\left[0,x\right],
\end{equation*}
is a strictly increasing continuous function with  $F(0)\leq 0$ and $F(x)\geq 0$  and then, there exists a unique  solution of \eqref{290}. 
$\Box$

%%%%%%%%%%%%%%%%%%%%%%%%%%%%%%%%%%%%%%%%%%%%%%%%%%%%%%%%%%%%%%%%%%%%%%%%%%%%%%
\prop \label{p12}
%%%%%%%%%%%%%%%%%%%%%%%%%%%%%%%%%%%%%%%%%%%%%%%%%%%%%%%%%%%%%%%%%%%%%%%%%%%%%%
In the case of the power utility $ U(x)  = \frac{x^p}{p},\,x>0,$ with $p<1$, $ p\neq 0,$ we suppose that the Assumptions \ref{a1}, \ref{a11}, \ref{a2}, \ref{a3}  hold,  $g(\xi)\in]0,x[$($\alpha$ -a.s.) and
$$\int_{\Xi}\left({\bf H}^{(q), *}_T(u)\right)^{1-p}d\alpha(u)<\infty .$$
Then,   the buyer's and seller's indifference prices  are  defined  respectively from the equations:
\begin{equation}
\int_{\Xi}[(1-\frac{p_T^b}{x}+\frac{g(u)}{x})^p-1]\left({\bf H}^{(q), *}_T(u)\right)^{1-p}d\alpha(u)=0
\label{30}
\end{equation} 
and
\begin{equation}
\int_{\Xi}[(1+\frac{p_T^s}{x}-\frac{g(u)}{x})^{p}-1]\left({\bf H}^{(q), *}_T(u)\right)^{1-p}d\alpha(u)=0
\label{300}
\end{equation} 
Moreover, the equations \eqref{30} and \eqref{300} have unique solutions belonging to the interval $[0,x]$. 
%%%%%%%%%%%%%%%%%%%%%%%%%%%%%%%%%%%%%%%%%%%%%%%%%%%%%%%%%%%%%%%%%%%%%%%%%%%

\rem  In the case when  $X(\xi)$ and $\xi$ are independent, the information quantity ${\bf H}^{(q), *}_T(u)$ does not depend on $u$ and  we get from Proposition \ref{p12} the following equations for indifference price:
\begin{equation*}
\int_{\Xi}[(1-\frac{p_T^b}{x}+\frac{g(u)}{x})^p-1]d\alpha(u)=0,
\end{equation*}
\begin{equation*}
\int_{\Xi}[(1+\frac{p_T^s}{x}-\frac{g(u)}{x})^{p}-1]d\alpha(u)=0.
\end{equation*} \rm

\pf\normalfont
The formula \eqref{30} follows from \eqref{73a} and \eqref{102}, then, the formula \eqref{300}  can be  obtained from the relation (\ref{pbps}).
\par We denote  for $ y\in\left[0,x\right]$
\begin{equation*}
F(y)=\int_{\Xi}\,\left[(1-\frac{y}{x}+\frac{g(u)}{x})^p -1\right]\,\left({\bf H}^{(q), *}_T(u)\right)^{1-p}d\alpha(u)
\end{equation*}
We see that $F$ is continuous strictly decreasing function for $p\in(0,1)$ on $[0,x]$ and that $F(0)\geq 0$ and $F(x)\leq 0$. Then the solution  of the equation exists by mean value theorem and it is unique. For $p<0$, $F$ is  a strictly increasing continuous function  with $F(0)\leq 0$ and $F(x)\geq 0$, then \eqref{30} has a unique solution. The case of seller's indifference price can be considered in a similar way.
$\Box$

%%%%%%%%%%%%%%%%%%%%%%%%%%%%%%%%%%%%%%%%%%%%%%%%%%%%%%%%%%%%%%%%%%%%%%%%%%%%%%
\prop \label{p13}
%%%%%%%%%%%%%%%%%%%%%%%%%%%%%%%%%%%%%%%%%%%%%%%%%%%%%%%%%%%%%%%%%%%%%%%%%%%%%%
In the case of the exponential utility  $U(x) = 1-e^{-\gamma x},\,x>0,$ with  $\gamma>0$ and under  the Assumptions \ref{a1}, \ref{a11}, \ref{a2}, \ref{a3} , the buyer's and seller's indifference prices 
verify: 
\begin{equation}
p_T^{b}=\frac{1}{\gamma} \ln\left[\frac{\int_{\Xi}\exp\bigg\{-{\bf{I}}(Q^{u, *}_T|P^{u}_T)\bigg\}d\alpha(u)}{\int_{\Xi} \exp\bigg\{-\gamma g(u)-{\bf{I}}(Q^{u, *}_T|P^{u}_T)\bigg\} d\alpha(u)}\right]
\label{31}
\end{equation} 
and
\begin{equation}
p_T^{s}=-\frac{1}{\gamma}\ln\left[ \frac{\int_{\Xi}\exp \bigg\{-{\bf{I}}(Q^{u, *}_T|P^{u}_T)\bigg\}d\alpha(u)}{\int_{\Xi} \exp\bigg\{\gamma g(u)-{\bf{I}}(Q^{u, *}_T|P^{u}_T)\bigg\}d\alpha(u)}\right]
\label{310}
\end{equation} 
%%%%%%%%%%%%%%%%%%%%%%%%%%%%%%%%%%%%%%%%%%%%%%%%%%%%%%%%%%%%%%%%%%%%%%%%%%%%%%

\rem  In the case when  $X(\xi)$ and $\xi$ are independent, the information quantity ${\bf{I}}(Q^{u, *}_T|P^{u}_T)$ does not depend on $u$ and  we get from Proposition \ref{p13} the following equations for indifference price:
\begin{equation*}
p_T^{b}= -\frac{1}{\gamma} \ln\left[\int_{\Xi} \exp\bigg\{-\gamma g(u)\bigg\} d\alpha(u)\right],
\end{equation*} 
\begin{equation*}
p_T^{s}=\frac{1}{\gamma}\ln\left[ {\int_{\Xi} \exp\bigg\{\gamma g(u)\bigg\}d\alpha(u)}\right].
\end{equation*} \rm

\pf\normalfont In the case of the exponential utility $I(y)=-\frac{1}{\gamma}(\ln y-\ln \gamma)$  and  from \eqref{102} and \eqref{74a}  we get the equation for buyer's indifference price:
\begin{equation*}
\int_{\Xi}\left(\lambda_g(u)-\lambda_0(u)\right)d\alpha (u)=0,
\end{equation*}
where  $\lambda_g$ is given by
$$
\lambda_g(u)=\gamma\exp\bigg\{-\gamma\left(x-p_T^b+g(u)\right)-{\bf{I}}(\,Q^{u, *}_T\,|_,P^{u*}_T\,)\bigg\},$$
and 
$$
\lambda_o(u)=\gamma\exp\bigg\{-\gamma\,x-{\bf{I}}(\,Q^{u, *}_T\,|_,P^{u*}_T\,)\bigg\}$$
These formulas give us \eqref{31}. The seller's indifference price \eqref{310} can be obtained from the relation \eqref{pbps}.
$\Box$

\subsection{Indifference prices and risk measure properties}

In this subsection we will show that indifference prices $p^s_T$ and $-p^b_T$
are risk measures. First we recall here the definition of risk measure.
\par The application $\rho : \mathcal F _T\rightarrow \mathbb R^+$ is convex risk measure if for all contingent claims $C_T^{(1)}, C_T^{(2)} \in \mathcal F_T$
and all $0<\gamma <1$ we have:
\begin{enumerate}
\item convexity of $\rho$ with respect to the claims:
$$\rho (\gamma\,C_T^{(1)} + (1-\gamma)\, C_T^{(2)})\leq \gamma \rho (C_T^{(1)}) +
(1-\gamma)\rho (C_T^{(2)})  $$
\item it is increasing function with respect to the claim:
$$\text{for}\,\,C_T^{(1)}\leq C_T^{(2)},\, \text{we have}\,\, \rho (C_T^{(1)}) \leq \rho (C_T^{(2)})$$
\item it is invariant with respect to the translation: for $m>0$
$$\rho (C_T^{(1)}+m)= \rho (C_T^{(1)})+m $$
\end{enumerate}
%%%%%%%%%%%%%%%%%%%%%%%%%%%%%%%%%%%%%%%%%%%%%%%%%%%%%%%%%%%%%%%%%%%%%%%%%%%%%%%%
\prop \label{p21}We suppose that The Assumptions \ref{a1}, \ref{a11}, \ref{a2}, \ref{a3}  hold. Then for HARA utilities  the indifference prices for sellers $p^s_T(g)$ and $(-p^b_T)$ for buyers obtained in the Propositions \ref{p11}, \ref{p12}, \ref{p13} are risk measures.
%%%%%%%%%%%%%%%%%%%%%%%%%%%%%%%%%%%%%%%%%%%%%%%%%%%%%%%%%%%%%%%%%%%%%%%%%%%%%%%%

\pf\normalfont 
We prove the claim for seller's indifference price since the corresponding properties  for $-p^b_T$ will follow from \eqref{pbps}.

\par (i) Let $U(x)= \ln (x), x>0$. From the Proposition \ref{p11} the indifference price for seller $p^s_T = p^s_T(g)$ is defined from the equation:
$$\int_{\Xi}\ln \left[1+\frac{p^{s}_T(g)}{x}-\frac{g(u)}{x}\right]d\alpha(u)=0.$$
Since for each $m\in \mathbb R^+$, $p^s_T(g+m)$ verify
$$\int_{\Xi}\ln \left[1+\frac{p^{s}_T(g+m)}{x}-\frac{g(u)+m}{x}\right]d\alpha(u)=0.$$
and the solution of this equation  is unique,
$$p^s_T(g+m) = p^s_T(g) + m$$
and, hence, the property (3) holds.
\par Let $g_1(u) \leq g_2(u)$ for $u\in \Xi$. Then we have:
$$ 0 = \int_{\Xi}\ln \left[1+\frac{p^{s}_T(g_1)}{x}-\frac{g_1(u)}{x}\right]d\alpha(u)\geq \int_{\Xi}\ln \left[1+\frac{p^{s}_T(g_1)}{x}-\frac{g_2(u)}{x}\right]d\alpha(u)$$
and it gives (2).
\par  We put $g(u) = \gamma \,g_1(u)+(1-\gamma)\,g_2(u) $. Then from concavity of $\ln$ we get:
$$\int_{\Xi}\ln \left[1+\frac{\gamma\,p^{s}_T(g_1)+ (1-\gamma)\,p^{s}_T(g_2) }{x}-\frac{\,g(u)}{x}\right]d\alpha(u)\geq $$
$$\gamma\,\displaystyle\int_{\Xi}\ln \left[1+\frac{p^{s}_T(g_1)}{x}-\frac{g_1(u)}{x}\right]d\alpha(u)
+ (1-\gamma)\, \int_{\Xi}\ln \left[1+\frac{p^{s}_T(g_2)}{x}-\frac{g_2(u)}{x}\right]d\alpha(u)$$
Then, since the right-hand side of previous expression is equal to zero,
$$p^s_T(\gamma \,g_1(u)+(1-\gamma)\,g_2(u))\leq \gamma\,p^{s}_T(g_1)+ (1-\gamma)\,p^{s}_T(g_2)$$
and we proved the relation (1).
\par (ii) Let $U(x) = \frac{x^p}{p}, p<1, p \neq 0$.  From the Proposition \ref{p12} the indifference price for seller is defined from the equation: 
$$\int_{\Xi}[(1+\frac{p_T^s(g)}{x}-\frac{g(u)}{x})^{p}-1]\left({\bf H}^{(q), *}_T(u)\right)^{1-p}d\alpha(u)=0$$
The properties (2) and (3) can be proved in the same way as in (i). Let us denote  by  $g(u) = \gamma \,g_1(u)+(1-\gamma)\,g_2(u) $ and let us suppose that $ 0<p<1$. Then using the
concavity of the function $(1+x)^p -1$ we have:
$$\int_{\Xi}[(1+\frac{\gamma\,p^{s}_T(g_1)+ (1-\gamma)\,p^{s}_T(g_2)}{x}-\frac{g(u)}{x})^{p}-1]\left({\bf H}^{(q), *}_T(u)\right)^{1-p}d\alpha(u)\geq $$
$\gamma \,\displaystyle\int_{\Xi}[(1+\frac{p_T^s(g_1)}{x}-\frac{g_1(u)}{x})^{p}-1]\left({\bf H}^{(q), *}_T(u)\right)^{1-p}d\alpha(u) +$\\
$$ \hspace{4cm}( 1-\gamma) \int_{\Xi}[(1+\frac{p_T^s(g_2)}{x}-\frac{g_2(u)}{x})^{p}-1]\left({\bf H}^{(q), *}_T(u)\right)^{1-p}d\alpha(u)$$
Since the right-hand side of above expression is equal to zero, we get the property (1). The case $p<0$ can be considered in similar way.
\par (iii) Let  $U(x) = 1-e^{-\gamma _0x},\,x>0,$ with  $\gamma _0>0$.  From the Proposition \ref{p13} the indifference price for the seller is defined by the formula:
$$p_T^{s}(g)=-\frac{1}{\gamma _0}\ln\left[ \frac{\int_{\Xi}\exp \bigg\{-{\bf{I}}(Q^{u, *}_T|P^{u}_T)\bigg\}d\alpha(u)}{\int_{\Xi} \exp\bigg\{\gamma _0g(u)-{\bf{I}}(Q^{u, *}_T|P^{u}_T)\bigg\}d\alpha(u)}\right]$$
We see directly from this formula that the properties (2) and (3) are verified.
Let us take $g(u) = \gamma \,g_1(u)+(1-\gamma)\,g_2(u)$. Then, by Holder inequality
with $p=\frac{1}{\gamma}$ and $q= \frac{1}{1-\gamma}$ we get:
$$\int_{\Xi} \exp\bigg\{\gamma _0g(u)-{\bf{I}}(Q^{u, *}_T|P^{u}_T)\bigg\}d\alpha(u)\leq
\left(\int_{\Xi} \exp\bigg\{\gamma _0g_1(u)-{\bf{I}}(Q^{u, *}_T|P^{u}_T)\bigg\}d\alpha(u)\right)^{\gamma}\,$$ $$\left(\int_{\Xi} \exp\bigg\{\gamma _0g_2(u)-{\bf{I}}(Q^{u, *}_T|P^{u}_T)\bigg\}d\alpha(u)\right)^{1-\gamma}$$
and it gives the property (1).
$\Box$

\end{section}

%%%%%%%%%%%%%%%%%%%%%%%%%%%%%%%%%%%%%%%%%%%%%%%%%%%%%%%%%%%%%%%%%%%%%%%%%%%%%%%
\begin{section}{Conditionally exponential Levy models}\label{s5}
%%%%%%%%%%%%%%%%%%%%%%%%%%%%%%%%%%%%%%%%%%%%%%%%%%%%%%%%%%%%%%%%%%%%%%%%%%%%%%%
\par We continue in the framework of Section 2. In addition, and it will be  specific for this part, we assume that conditionally to $\xi=u$, $X(\xi)$ is a Levy process. It means that $P^u$ is the law of Levy process with the parameters $(b^u, (\sigma^u)^2, \nu^u),$ where $b^u $ is a drift parameter, $(\sigma^u)^2$ is a parameter related with a continuous martingale part and $\nu^u$ is a Levy measure, such that 
$$\int_{\mathbb{R}}(x^2\wedge 1 )\,\nu^u(dx)<\infty$$ Since $P^u\ll P$ we deduce that $(\sigma^u)^2$ does not depend on $u$ and then $(\sigma^u)^2=\sigma^2$. We recall that  according to the Levy-Ito decomposition theorem, conditional Levy process $X$ with the generating  triplet $(b^u,\sigma^2,\nu^u)$  can be represented in the following form
\begin{equation*}
X_t=\sigma W^{u}_t+b^{u}t+\int_0^t\int_{|x|>1}xN^{u}(ds, dx)+\int_0^t\int_{|x|\leq 1}x\tilde{N}^{u}(ds, dx),
\end{equation*}
where $W^{u}$ is a $(P^u,{\bf F})$ standard Wiener process, $N^{u}(ds, dx)$ is a $(P^u,{\bf F})$ Poisson random measure and $\tilde{N}^{u}(ds, dx)$ is a $(P^u,{\bf F})$ compensated Poisson random measure with a compensator $\nu^{u}(dx)ds$.
\par The parameters of Levy process define entirely the law of the process via its  one-dimensional distributions : for all $\lambda\in\mathbb{R}$
\begin{equation*}
E_{P^u}e^{i\lambda X_t}=e^{t\psi^u(\lambda)},
\end{equation*}
where the characteristic exponent $\psi^u(\lambda)$ is given by Levy-Kinchin formula:
\begin{equation*}
\psi^u(\lambda)=i\lambda b^u-\frac{1}{2}\lambda^2\sigma^2+ \int_{\mathbb{R}}\left(e^{itx}-1-x1_{\{|x|\leq 1\}}\right)\nu^u(dx).
\end{equation*}
\par Exponential Levy models was very well studied (see for instance \cite{A}, \cite{Sa}). In particular, the notion  of minimal entropy martingale measure was introduced first in \cite{M}, the question of existence of $f$-divergences minimal martingale measures, for the first time was studied in  \cite{GR}, and for classical $f$-divergences for the exponential Levy models it was done in its generalised version in \cite{CV1}.
Again, we denote by $\beta^{u, *}$ and $Y^{u, *}(x)$ two $(P^u,{\bf F})$-predictable processes known as Girsanov parameters for the changing of measure from $P^u$ to $Q^{u, *}$ such that: $\forall t\geq 0$ and $P^u$-a.s.
\begin{equation*}
\int_0^t\int_{\mathbb{R}}|l(x)\,(Y_s^{u, *}(x)-1)|\,\nu^u(dx)\,ds <\infty,\ \ \int_0^t(\beta_s^{u, *})^2\,ds<\infty.
\end{equation*}
 
We recall that for HARA utilities, the equivalent $f$-divergence minimal martingale measures when they exist, have Levy preservation property, i.e. being Levy process under $P^u$, the process remains Levy process under $Q^{u, *}$.  More about preservation of Levy property see \cite{CV2}. The preservation of Levy property implies that all information processes  introduced in  section \ref{s4} are deterministic and this fact  simplifies very much the expression for maximal utility of Theorem \ref{t5}.
%%%%%%%%%%%%%%%%%%%%%%%%%%%%%%%%%%%%%%%%%%%%%%%%%%%%%%%%%%%%%%%%%%%%%%%%%%%%
\prop \label{p14}
%%%%%%%%%%%%%%%%%%%%%%%%%%%%%%%%%%%%%%%%%%%%%%%%%%%%%%%%%%%%%%%%%%%%%%%%%%%%
Let $U(x)=\ln x,\,x>0,$ and Assumption \ref{a1} holds. We suppose that there exists a solution $\beta^{u}$ to the equation 
\begin{equation}
b^{u}+\beta^{u}\sigma^2+\int_{\mathbb{R}}\left(Y^{u}(x)-1\right)1_{\{|x|\leq 1\}}\nu^{u}(dx)=0,
\label{91}
\end{equation}
with 
\begin{equation}\label{p14y}
Y^{u}(x)=(1-\beta^{u}\,x\,1_{\{|x|\leq 1\}})^{-1}
\end{equation}
and such that $Y^{u}(x)>0 \ \ \nu^{u}-$a.s. Then,  there exists $f$-divergence minimal equivalent  martingale measure $Q^{u}$ and the  corresponding  information process is equal to:
\begin{equation}
\mathcal{I}_T(u)=T\bigg\{\frac{1}{2}\left(\beta^{u}\sigma\right)^2+\int_{\mathbb{R}}\left(-\ln\left(Y^{u}(x)\right)+Y^{u}(x) -1\right)\nu^u(dx)\bigg\}
\label{112}
\end{equation}
If we assume, in addition, that $g(\xi)\in ]0,x[$ ($\alpha $-a.s.) and that $\ln g(\xi)$, $\ln (x-g(\xi))$ and $\mathcal{I}_T(\xi)$ are $\alpha$-integrable random variables, then
\begin{equation}
V(x,0)=\int_{\Xi}\mathcal{I}_T(u)d\alpha(u)-\ln x
\label{113}
\end{equation}
and for the buyer of option 
\begin{equation}
V(x-p_T^b,g)=V(x,0)+\int_{\Xi}\ln\left(1-\frac{p_T^b}{x}+\frac{g(u)}{x}\right)d\alpha(u),
\label{114}
\end{equation}
for the seller of option
\begin{equation}
V(x+p_T^s,g)=V(x,0)+\int_{\Xi}\ln\left(1+\frac{p_T^s}{x}-\frac{g(u)}{x}\right)d\alpha(u).
\label{115}
\end{equation}
Moreover, the indifference prices $p_T^b$, $p_T^s$ are defined by the formulas \eqref{29} and \eqref{290} respectively.
%%%%%%%%%%%%%%%%%%%%%%%%%%%%%%%%%%%%%%%%%%%%%%%%%%%%%%%%%%%%%%%%%%%%%%%%%%%%%%

\pf\normalfont
The stochastic exponent of $X$ will be a $(Q^u,{\bf F})$ local martingale  if the process $X$  is a $(Q^u,{\bf F})$ local martingale. The process $X$ will be a local martingale under the measure $Q^u$ if and only if the corresponding drift parameter $B^{Q^u}$ is equal to 0 for each $t$. It was shown in \cite{Kl} that Girsanov parameters of the minimal martingale measure does not depend on $(\omega ,t)$ (see also \cite{CV2}). Since  $\beta^{u}$ and $Y^u(x)$ denote the Girsanov parameters for the changing of measure from $P^u$ to $Q^u$, according to \cite{JSh}, Theorem 3.24, p. 159, we have $\forall t\in\left[0,T\right]$
\begin{equation*}
B^{Q^u}_t=b^u t+\beta^u\sigma^2 t+t\int_{\mathbb{R}}(Y^u(x)-1)1_{\{|x|\leq 1\}}(x)d\nu(x).
\end{equation*}
Equating $B^{Q^u}$ to 0, one gets the relation \eqref{91}. 
\par It was shown in \cite{Kl} (see also \cite{CV2}) that if $S_t= \exp (\tilde{X}_t)$ then 
$$Y^u(\Delta \tilde{X}) = (1- \beta ^u (e^ {\Delta \tilde{X}}-1))^{-1}$$
But at the same time $S_t= \mathcal E (X)_t$ and writing the relation for the jumps we get
\begin{equation}\label{jump}
\exp (\Delta \tilde{X})-1= \Delta X_t 1_{\{|\Delta X_t|\leq 1\}}
\end{equation}
and it gives the formula \eqref{p14y}.
From \eqref{71} we get the expression of the corresponding information process \eqref{112}. From Theorem \ref{t5} we obtain the expressions for maximal expected utility, and Proposition \ref{p11} gives us the formulas for indifference prices.
$\Box$

%%%%%%%%%%%%%%%%%%%%%%%%%%%%%%%%%%%%%%%%%%%%%%%%%%%%%%%%%%%%%%%%%%%%%%%%%%%%%%
\prop \label{p15}
%%%%%%%%%%%%%%%%%%%%%%%%%%%%%%%%%%%%%%%%%%%%%%%%%%%%%%%%%%%%%%%%%%%%%%%%%%%%%
Let $U(x)=\frac{x^p}{x},\ \ x>0,$ with $ p<1, p\neq 0$ and Assumption \ref{a1} holds. We suppose that there exists a solution to the equation \eqref{91} with 
\begin{equation}
\label{p15y}
Y^u(x)=\left(1-\frac{|p|}{(p-1)^2}\beta^{u}\,x\,1_{\{|x|\leq 1\}}\right)^{p-1}
\end{equation}
such that $1-\displaystyle\frac{|p|}{(p-1)^2}\beta^{u}\,x\,1_{\{|x|\leq 1\}}>0$ ($\nu^{u}-$a.s.) Then, there exists $f$-divergence minimal equivalent martingale measure $Q^u$ and the corresponding Hellinger type process of order $q=\frac{p}{p-1}$ is given by:
\begin{multline}\label{p150}
h^{(q)}_T(u)=T\bigg\{\frac{1}{2}q(q-1)\left(\beta^{u }\sigma\right)^2+\int_{\mathbb{R}}\left[\left(Y^{u}(x)\right)^{q}\right.
\left.-q (Y^{u}(x)-1)-1\right]\nu ^u(dx)\bigg\}
\end{multline}
If we assume, in addition, that  $g(\xi)\in ]0,x[$ ($\alpha $-a.s.) and that 
$e^{h^{(q)}_T(\xi)}$ is $\alpha$-integrable random variable, then
\begin{equation*}
V(x,0)=x^p \int_{\Xi}e^{h^{(q)}_T(u)}d\alpha(u),
\end{equation*}
for the buyer of the option 
\begin{equation*}
V(x-p_T^b,g)=\int_{\Xi}(x-p_T^b+g(u))^p\,e^{h^{(q)}_T(u)}d\alpha(u),
\end{equation*}
for the seller of the option 
\begin{equation*}
V(x+p_T^s,g)=\int_{\Xi}(x+p_T^b-g(u))^p\,e^{h^{(q)}_T(u)}d\alpha(u),
\end{equation*}
Moreover,  the buyer's and seller's indifference prices are defined by the formulas \eqref{30} and \eqref{300} respectively with ${\bf H}^{(q),*}_T(u) = e^{h^{(q)}_T(u)}$.

\pf\normalfont
The same reasons as in the proof of the Proposition \ref{p14} gives us \eqref{91}. It was shown in \cite{JKM} that Girsanov parameters of the minimal martingale measure does not depend on $(\omega ,t)$ (see also \cite{CV2}) and that if $S_t= \exp (\tilde{X}_t)$ then 
$$Y^u(\Delta \tilde{X}) = \left(1- \frac{|p|}{(p-1)^2}\beta ^u (e^ {\Delta \tilde{X}}-1)\right)^{p-1}$$
But then $S_t= \mathcal E (X)_t$ and \eqref{jump}
 gives us the formula \eqref{p15y}.
Then from \eqref{88} we deduce the expression \eqref{p150}. From Theorem \ref{t5} we obtain the expressions for maximal expected utility, and Proposition \ref{p12} gives us the formulas for indifference prices.
$\Box$
%%%%%%%%%%%%%%%%%%%%%%%%%%%%%%%%%%%%%%%%%%%%%%%%%%%%%%%%%%%%%%%%%%%%%%%%%%%%%%
\prop \label{p16}
%%%%%%%%%%%%%%%%%%%%%%%%%%%%%%%%%%%%%%%%%%%%%%%%%%%%%%%%%%%%%%%%%%%%%%%%%%%%%% 
Let $U(x)=1-e^{-\gamma x},$ $x>0,$ with $\gamma>0$ and  Assumption \ref{a1} holds. We suppose that there exists a solution to the equation \eqref{91} 
with 
\begin{equation}\label{p16y}
Y^{u}(x)=\exp\{\beta^{u } x\,1_{\{|x|\leq 1\}}\}
\end{equation}
Then, there exists $f$-divergence minimal equivalent martingale measure $Q^u$ and the corresponding information process  is given by:
\begin{equation}\label{p160}
I_T(u)=T\bigg\{\frac{1}{2}\left(\beta^{u}\sigma\right)^2+\int_{\mathbb{R}}\left[Y^{u}(x)\ln Y^{u}(x)-Y^{u}(x)+1\right]\nu^u(dx)\bigg\},
\end{equation}
If we suppose that $I_T(u)$ is finite ($\alpha$-a.s.) then we have:
\begin{equation*}
V(x,0)=1-\int_{\Xi}\exp \{-\gamma x-I_T(u)\}\,d\alpha(u),
\end{equation*}
for the buyer, 
\begin{equation*}
V(x-p_T^b,g)=1-\int_{\Xi}\exp \{-\gamma \left(x-p_T^b+g(u)\right)-I_T(u)\}\,d\alpha(u),
\end{equation*}
for the seller,
\begin{equation*}
V(x+p_T^s,g)=1-\int_{\Xi}\exp \{-\gamma \left(x+p_T^s-g(u)\right)-I_T(u)\}\,d\alpha(u).
\end{equation*}
Moreover,  the buyer's and seller's indifference prices are defined by the formulas \eqref{31} and \eqref{310} respectively with ${\bf I}(Q^{u,*}_T\,|\, P^u_T)  = I_T(u)$.

\pf\normalfont
It was shown in \cite{ES} that Girsanov parameters of the minimal martingale measure does not depend on $(\omega ,t)$ (see also \cite{CV2}). Since  $\beta^{u}$ and $Y^u(x)$ denote the Girsanov parameters for the changing of measure from $P^u$ to $Q^u$, according to \cite{JSh}, Theorem 3.24, p. 159, we have  \eqref{91}. 
\par It was shown in \cite{HS} (see also \cite{CV2})that if $S_t= \exp (\tilde{X}_t)$ then 
$$Y^u(\Delta \tilde{X}) = e^ {\beta ^u\,(e^{\Delta \tilde{X}}-1)}$$
But since $S_t= \mathcal E (X)_t$  and \eqref{jump}
we get the formula \eqref{p16y}.
From \eqref{83} we get the expression of the corresponding information process \eqref{p160}. From Theorem \ref{t5} we obtain the expressions for maximal expected utility, and Proposition \ref{p13} gives us the formulas for indifference prices.
$\Box$
\end{section}

%%%%%%%%%%%%%%%%%%%%%%%%%%%%%%%%%%%%%%%%%%%%%%%%%%%%%%%%%%%%%%%%%%%%%%%%%%%%%%
\begin{section}{Applications to Geometric Brownian motion case}\label{s6}
%%%%%%%%%%%%%%%%%%%%%%%%%%%%%%%%%%%%%%%%%%%%%%%%%%%%%%%%%%%%%%%%%%%%%%%%%%%%%%%

\par Let $(W^{(1)}, W^{(2)})$ bi-dimensional standard Brownian motions with correlation $\rho$, $|\rho| <1$ on $[0,T]$.  Let $\mu _1,\mu _2 \in \mathbb R$ and $\sigma _1>0, \sigma _2>0$. We put
$$S^{(1)}_t= \exp \{(\mu _1 - \frac{\sigma^2_1}{2})t + \sigma _1\,W^{(1)}_t \}$$
$$S^{(2)}_t= \exp \{(\mu _2 - \frac{\sigma^2_2}{2})t + \sigma _2\,W^{(2)}_t \} $$
for two risky assets. 
\par The first asset will play the role of $S(\xi)$ and $X_t(\xi)
= \mu _1\,t + \sigma _1\,W^{(1)}_t$ in this case. We take $\xi = W^{(2)}_{T'}$ instead of $S^{(2)}_{T'}$ since they generate the same $\sigma$-algebras. In this case $\alpha = \mathcal N(0,T')$.
\par We know that for all $t\in [0,T]$
$$W^{(1)}_t = \rho W^{(2)}_t +\sqrt{1-\rho^2} \gamma_t$$
where $\gamma$ is independent from $W^{(2)}$ standard Brownian motion. Then, the conditional law of $X$ given $W_{T'}^{(2)}=u$ coincide with the law of 
\begin{equation}\label{e71}
X_t(u) = \mu _1 t +\sigma _1 \rho V_t(u) + \sigma _1 \sqrt{1-\rho^2} \gamma_t
\end{equation}
where $V(u)$ is a Brownian bridge starting from 0 at $t=0$ and ending in $u$ at $t=T'$ which is independent from the process $\gamma$. As known,
$$ V_t(u) = \int _0^T \frac{u-V_s(u)}{T' -s} ds +\eta _t$$
where $\eta$ is standard Brownian motion independent from $\gamma$. Finally, since
$\hat{\gamma}= \rho \eta +\sqrt{1-\rho^2}\gamma$ is again standard Brownian motion,
we get:
\begin{equation}\label{e72}
X_t(u) = \mu _1 t +\sigma _1\rho \int _0^t \frac{u-V_s(u)}{T' -s} ds + \sigma _1 \,\hat{\gamma_t}
\end{equation}
Hence, $P^u_t <\!\!< P_t$ for all $u\in\mathbb R$ and $t\in [0,T]$, and the Assumptions
\ref{a1} and \ref{a11} are satisfied.
\par Let au calculate the conditional law $\alpha^t = P(\xi\,|\, \mathcal F_t)$
given $\mathcal F_t= \sigma (W^{(1)}_s, s\leq ~t)$. By Markov property we get: for $A\in \mathcal B(\mathbb R)$
$$\alpha ^t(A) = P( W^{(2)}_{T'}\in A\,|\,\mathcal F_t) = P( W^{(2)}_{T'}\in A\,|\,W^{(1)}_t)= P( W^{(2)}_{T'} - W^{(2)}_t +  W^{(2)}_t \in A\,|\,W^{(1)}_t)$$
Since $ W^{(2)}_{T'} - W^{(2)}_t$ is independent from $(W^{(1)}_t, W^{(2)}_t)$, the law of $\xi$ given $W^{(1)}_t=x$ is $\mathcal N(\rho\,x, T'-\rho ^2t)$. So, since $T' - \rho^2t \neq 0$ for $t\in [0,T]$, it is equivalent to the law of $W^{(2)}_{T'}$
being $\mathcal N (0, T')$.
\par To give the formulas for indifference price it is convenient to remark that $Q^{u,*}$ is a unique martingale measure which annulate  the drift of $X(u)$
given by
$$B_t(u) = \mu _1 t +\sigma _1\rho \int _0^t \frac{u-V_s(u)}{T' -s} ds $$
If we denote
$$ \beta^u_s=  \mu _1 +\sigma _1\rho \frac{u-V_s(u)}{T' -s} $$
then
$$\frac{dQ^{u,*}_T}{dP^u_T}= \exp \{ \sigma _1\int _0^T \beta^u_s d \hat{\gamma}_s + \frac{\sigma _1^2}{2} \int _0^T (\beta^u_s)^2ds\}$$
\par Let us write the information processes corresponding to $( P^u, Q^{u,*})$.
For Hellinger process we have:
$$ h_t^{(q)}= \frac{q (1-q)\,\sigma _1^2}{2} \int _0^T (\beta^u_s)^2ds$$
and 
$${\bf H}_T^{(q)}(u) = E_{P^u}\left[ \exp\{ \frac{q (1-q)\,\sigma _1^2 }{2} \int _0^T (\beta^u_s)^2ds\}\right].$$
For Kullback-Leibler  process we get:
$$ I^*_T(u)= \frac{\sigma _1^2}{2} \int _0^T (\beta^u_s)^2ds$$
and  Kullback-Leibler information
$${\bf I} ( Q^{u,*}\,|\, P^u) = E_{Q^{u,*}} (\,I^*_T(u)\,).$$
For the entropy of $P^u_T$  with respect to $Q^{u,*}$ we deduce that:
$$ \mathcal{I}_T^*(u) = \frac{\sigma _1^2}{2} \int _0^T (\beta^u_s)^2ds$$
and 
$${\bf I} ( P^u\,|\, Q^{u,*}) = E_{P^u} ( \mathcal{I}^*_T(u)).$$
%%%%%%%%%%%%%%%%%%%%%%%%%%%%%%%%%%%%%%%%%%%%%%%%%%%%%%%%%%%%%
\prop \label{p71}For mentioned three information quantities we have the following result:
$${\bf I} ( P^u\,|\, Q^{u,*})= \frac{\sigma _1^2}{2}\left[\left( \mu _1 - \frac{\sigma _1 \rho u}{T'}\right)^2\,T + \frac{\sigma_1^2 \rho ^2}{T'}\left( T' \ln(\frac{T'}{T'-T})-T\right)\right],$$
$${\bf I} ( Q^{u,*}\,|\, P^u)= \frac{\sigma _1^2}{2} \left\{ \mu _1^2\,T + 2 \sigma _1\,\mu _1\,\rho\,u\,\ln(\frac{T'}{T'-T}) + \sigma_1^2 \rho ^2\,u^2 \frac{T}{T'(T'-T)}\right.$$
$$\left.+ \sigma_1^2 \rho ^2\left[\frac{T}{T'-T}-\ln(\frac{T'}{T'-T})\right] \right\},$$

$${\bf H}_T^{(q)}(u)= \left(\frac{T'}{T'-T+qT}\right)^{1/2}\exp \left\{-\frac{(1-q)}{2}\left[\frac{u^2}{T'}- \frac{(u+cT)^2}{T'-T+qT}\right] \right\}$$
where  $q> -(\displaystyle\frac{T'}{T}-1)$ and $c= \displaystyle\frac{\mu _1}{\sigma _1\,\sqrt{1-\rho ^2}}$\\
%%%%%%%%%%%%%%%%%%%%%%%%%%%%%%%%%%%%%%%%%%%%%%%%%%%%%%%%%%%%%
\pf\normalfont We begin with the calculus of $ E_{P^u} (\beta^u_t)^2$. We have:
$$(\beta^u_t)^2 = \mu_1^2 + 2\mu _1 \sigma _1 \rho \frac{u-V_t(u)}{T' -t}+
\sigma _1^2\rho ^2\frac{(u-V_t(u))^2}{(T' -t)^2}$$
From second representation for Brownian bridge we know that
$$(V_t(u))_{0\leq t\leq T'} \stackrel{\mathcal L}{=} 
(W_t - \frac{t}{T'}(W_{T'}-u))_{0\leq t\leq T'}$$
Then,
$$ E_{P^u}(u-V_t(u))= \frac{u(T'-t)}{T'}$$
and
$$E_{P^u}(u-V_t(u))^2 = \frac{t(T'-t)}{T'}+ u^2\frac{(T'-t)^2}{(T')^2}$$
Hence,
$$E_{P^u} (\beta^u_t)^2= (\mu _1+ \frac{\sigma _1 \rho u}{T'})^2 + \sigma _1^2 \rho ^2\frac{t}{T'(T'-t)},$$
and using Fubini theorem and the expression for $\mathcal{I}_T^*(u)$ we get the first equality.
\par The semi-martingale characteristics $X(u)$ under $P^u$ are: $( B(u), I, 0)$ where $I(t)=t$ and $$B_t(u)=\mu _1 t +\sigma _1\rho \int _0^t \frac{u-V_s(u)}{T' -s} ds.$$ From another side, the change of the measure  $P^u$ into $Q^{u,*}$ annulate the drift of $X(u)$, i. e. the semi-martingale
characteristics  of $X(u)$ will be $(0, I, 0)$. One of the possibilities to do this
is annulate the drift of $V(u)$ transforming this process into Brownian motion, then to annulate the drift $\mu _1 I$ using independent Brownian motion $\gamma$.
All successive equivalent change of the measures will give the same final result in terms of information quantities.
Hence,
$$E_{Q^{u,*}}(\beta^u_t)^2 = \mu _1^2 + 2\mu _1 \sigma _1 \rho \frac{u}{T' -t} + 
\sigma _1^2 \rho ^2 \frac{u^2 +t}{(T'-t)^2}$$ 
Using Fubini theorem and the expression for $I_T(u)$ given previously, we get the second result.
\par Now, we calculate ${\bf H}_T^{(q),*}$ applying the definition of Hellinger integral, namely,
\begin{equation}\label{f00}
{\bf H}_T^{(q),*} = E_{P^u} (Z^*_T(u))^q = E_{Q^{u,*}}(Z^*_T(u))^{q-1}
\end{equation}
We will find $Z^*_T(u)$ first. For that we remark that $P^u$ is the law of the process $X(u)=(X_t(u))_{0\leq t\leq T}$ with 
$$X_t(u) = \mu _1 t +\sigma _1 \rho V_t(u) + \sigma _1 \sqrt{1-\rho^2} \gamma_t$$
where $V(u)$ is Brownian bridge independent from standard Brownian motion $\gamma$.
As it was mentioned, the change of the measure $P^u$ into $Q^{u,*}$ annulate the drift of $X(u)$. This annulation can be made in two steps: annulate the drift of $V(u)$ transforming $V(u)$ into standard Brownian motion, and then, annulate the drift $\mu _1 I$ using the change of the measure related with the process $\gamma$.
More precisely, we do the following transformations:
$$(V(u), \sigma _1\sqrt{1-\rho ^2} \gamma + \mu _1 I)\rightarrow (W, \sigma _1\sqrt{1-\rho ^2} \gamma + \mu _1 I)\rightarrow (W, \sigma _1\sqrt{1-\rho ^2} \gamma )$$
where $W$ and $\gamma$ are standard independent Brownian motions.
\par Let us denote $\hat{P}^u$ the law of $(V_t(u))_{0\leq t\leq T}$ and by $P_T$ the law of $(W_t)_{0\leq t\leq T}$. We show that
\begin{equation}\label{f0}
Z_T^{(1)} = \frac{dP_T}{d\hat{P}_T}= \sqrt{\frac{T'}{T'-T}}\exp \left\{-\frac{(u-W_T)^2}{2(T'-T)}+ \frac{u^2}{2T'} \right\}
\end{equation}
In fact, for any measurable bounded functionals $F$ and $G$ we have:
\begin{equation}\label{f1}
E[F(W_s, s\leq T) G( W_{T'})] = E\left\{ E(F(W_s, s\leq T)\,|\, W_T)\int _{\mathbb R}
\displaystyle\frac{\exp (\frac{(x-W_T)^2}{2(T'-T)})}{\sqrt{2\pi(T'-T)}}G(x) dx\right\}
\end{equation}
since $W_{T'} = W_{T'}-W_T + W_T \stackrel{\mathcal L}{=} \tilde{W}_{T'-T} + W_T$
where $\tilde{W}$ is independent from $W$ standard Brownian motion. Hence,
$$E[F(W_s, s\leq T) G( W_{T'})] = E\left[F(W_s, s\leq T)\int _{\mathbb R}
\frac{\exp (\frac{(x-W_T)^2}{2(T'-T)})}{\sqrt{2\pi(T'-T)}}G(x) dx\right]$$
In addition,
\begin{equation}\label{f2}
E[F(W_s, s\leq T) G( W_{T'})] = E\left[ E(F(W_s, s\leq T)\,|\, W_{T'})\int _{\mathbb R}\frac{\exp (\frac{x^2}{2T'})}{\sqrt{2\pi T'}}G(x)dx\right]
\end{equation}
Since \eqref{f1} and \eqref{f2} are verified for any bounded $G$ we get:
$$E(F(W_s, s\leq T)\,|\, W_{T'}) = E\left[\,F(W_s, s\leq T)\frac{\exp (\frac{(x-W_T)^2}{2(T'-T)}+ \frac{x^2}{2T'}}{\sqrt{(T'-T)/T'}}\,\right]$$
The last equality proves that \eqref{f0} holds.
\par To annulate the drift $\mu _1 I$ we use  the process $\gamma$ and the change of the measure with the density:
$$Z_T^{(2)} = \exp\left\{ \frac{\mu _1\,\gamma _T }{\sigma _1 \sqrt{1-\rho ^2}}- 
\frac{\mu _1^2\, T}{2\sigma _1^2 (1-\rho ^2)}\right\}$$
So, the measure  which transform $(V(u), \sigma _1\sqrt{1-\rho ^2} \gamma + \mu _1 I)$ into $(W, \sigma _1\sqrt{1-\rho ^2} \gamma]$ has a density $Z_T^{(1)}Z_T^{(2)}$.
Using the theorem about of change of variables we find equivalent to $Z^*_T(u)$ expression in law with respect to $Q^{u,*}$:
$$Z^*_T(u) \stackrel{\mathcal L }{= }\int Z_T^{(1)}(W_T-y) Z_T^{(2)}(y) dP_{\gamma _T}(y)$$
Simple calculations gives us that
$$Z^*_T(u) \stackrel{\mathcal L }{= } \exp \left(\frac{ u^2 -(u-W_T +c T)^2}{2T'}\right)$$
with $c=\frac{\mu _1}{\sigma _1\,\sqrt{1-\rho ^2}}$. Then, using \eqref{f00}
we get after simple calculus the third result.$\Box$

Now, from Propositions \ref{p11}, \ref{p12}, \ref{p13} and \ref{p71} we can
find indifference prices taking $\alpha = \mathcal{N}(0, T')$.
\end{section}

\noindent \section{Acknowledgements} \rm  This work is supported in part by  ANR-09-BLAN-0084-01 of the Department of Mathematics of Angers's University.

\end{document}